\documentclass[aps,prc,floatfix,showpacs,twocolumn]{revtex4-1}
\usepackage{ulem}
\usepackage{bm}
\usepackage{color}
\usepackage{amssymb}
\usepackage{amsmath}
\usepackage{graphicx}
\usepackage{amsfonts}
\usepackage{slashed}
\usepackage{pstricks}
\usepackage{float}
\usepackage{hyperref}
\usepackage{array}
\allowdisplaybreaks
\usepackage{dcolumn}
\usepackage{epsf}
\usepackage{tabularx}
\usepackage{soul}
\usepackage{wrapfig}

\begin{document}
\title{Benchmark calculations of pure neutron matter with realistic nucleon-nucleon interactions}
\author{
M.\ Piarulli$^{\, {\rm a} }$,
I.\ Bombaci$^{\, {\rm b,c} }$, 
D.\ Logoteta$^{\, {\rm b,c} }$,
A.\ Lovato $^{\, {\rm d,e} }$, and
R.\ B. \ Wiringa$^{\, {\rm d} }$,
}
\affiliation{
$^{\,{\rm a}}$\mbox{Physics Department, Washington University, St Louis, MO 63130, USA}\\
$^{\,{\rm b}}$\mbox{Dipartimento di Fisica ``E. Fermi'', Universit\`a di Pisa, Largo B. Pontecorvo 3, I-56127 Pisa, Italy}\\
$^{\,{\rm c}}$\mbox{INFN, Sezione di Pisa, Largo B. Pontecorvo 3, I-56127 Pisa, Italy}\\
$^{\,{\rm d}}$\mbox{Physics Division, Argonne National Laboratory, Argonne, IL 60439, USA}\\
$^{\,{\rm e}}$\mbox{INFN-TIFPA Trento Institute of Fundamental Physics and Applications, 38123 Trento, Italy}
}
\date{\today}

\begin{abstract}
{We report benchmark calculations of the energy per particle of pure neutron matter as a function of the baryon density using three independent many-body methods: Brueckner-Bethe-Goldstone, Fermi hypernetted chain/single-operator chain, and auxiliary-field diffusion Monte Carlo.  Significant technical improvements are implemented in the latter two methods.  The calculations are made for two distinct families of realistic coordinate-space nucleon-nucleon potentials fit to scattering data, including the standard Argonne $v_{18}$ interaction and two of its simplified versions, and four of the new Norfolk $\Delta$-full chiral effective field theory potentials.  The results up to twice nuclear matter saturation density show some divergence among the methods, but improved agreement compared to earlier work.  We find that the potentials fit to higher-energy nucleon-nucleon scattering data exhibit a much smaller spread of energies.
}
\end{abstract}

\maketitle

\section{Introduction} 
The quest for understanding static and dynamic properties of nuclear systems in terms of nucleon-nucleon ($NN$) and three-nucleon ($3N$) forces, and consistent electroweak currents has long been considered one of the most challenging effort of nuclear theory. Over the past twenty years, establishing this {\it basic model} of nuclear physics has undergone substantial progress, driven by two major factors. First, since the advent of chiral effective field theory ($\chi$EFT), originally proposed by Weinberg in the early 1990's~\cite{Weinberg:1990rz,Weinberg:1991um}, we can now systematically develop nuclear many-body interactions~\cite{Epelbaum:2008ga,Machleidt:2011zz,Reinert:2017usi,Entem:2017gor} and consistent electroweak currents~\cite{Pastore:2009is,Kolling:2009iq,Pastore:2011ip,Kolling:2011mt,Baroni:2015uza,Krebs:2016rqz,Baroni:2016xll} that are rooted in the fundamental symmetries exhibited by the underlying theory of Quantum Chromodynamics. Second, present computational resources allow to employ these interactions and current in sophisticated many-body methods to compute a variety of nuclear systems with controlled approximations~\cite{Schulze:1995zz,bbg2,Carbone:2013eqa,Barrett:2013nh,Hagen:2013nca,Carlson:2014vla,Hergert:2015awm}. The chief challenge for the {\it basic model} is to accurately describe properties of atomic nuclei -- including their spectra, form factors, transitions, low-energy scattering, and response -- while simultaneously predicting properties of infinite matter, e.g., pure neutron matter (PNM), relevant to the structure and internal composition of neutron stars. 

The last few years has marked the birth of the multi-messenger astronomy era~\cite{GBM:2017lvd}, which has opened new windows to probe the constituents of matter and their interactions under extreme conditions that cannot be reproduced in terrestrial laboratories. The first direct detection of gravitational-waves from coalescing neutron stars by the LIGO-Virgo interferometer network~\cite{ligo_web,virgo_web}, followed by a short burst of $\gamma$ rays and later optical and infrared signals -- the event GW170817~\cite{TheLIGOScientific:2017qsa,GBM:2017lvd} -- effectively constrain their masses, spin and tidal deformability~\cite{Abbott:2018wiz,Tews:2018chv,Fasano:2019zwm}. In addition, the multiple measurements of two-solar masses neutron stars~\cite{Demorest:2010bx,Lynch:2012vv,Antoniadis:2013pzd,Arzoumanian:2017puf,Cromartie:2019kug} are posing intriguing questions about how dense matter can support such large masses against gravitational collapse.

The equation of state (EoS) of strongly interacting matter is a thermodynamic relation between the energy (pressure), the baryon density, and the temperature. Whilst the description of core-collapse supernovae and the formation and cooling of proto-neutron stars requires finite-temperature EoS, already a few minutes after its birth, neutron star properties can be safely described using the EoS of cold (zero temperature) neutron-rich matter~\cite{Prakash:1996xs}. In the region between the inner crust and the outer core ($\sim 0.5-2\rho_0$, with $\rho_0=0.16$\,fm$^{-3}$ being the nuclear saturation density), neutron stars are mainly comprised of neutrons, in $\beta$ equilibrium with a small fraction of protons, electrons and muons. Different scenarios have been suggested to model the high-density regime, from nucleon degrees of freedom only but with many-nucleon forces and relativistic effects~\cite{Friedman:1981qw,WFF88,Akmal:1998cf,Lynn:2015jua,Bombaci:2018ksa}, to including the formation of heavier baryons containing strange quarks~\cite{Glendenning:1984jr,Vidana:2010ip,Lonardoni:2014bwa,Chatterjee:2015pua,Haidenbauer:2016vfq}, to quark matter~\cite{Ranea-Sandoval:2015ldr,Alford:2015dpa,Mariani:2016pcx,Bombaci:2016xuj}, or other more exotic condensates~\cite{Pandharipande:1995qf,Glendenning:1997ak,Mukherjee:2008un}. While the determination of the maximum mass of a neutron star requires knowing the EoS up to several times nuclear saturation density, the EoS up to $\sim 2 \rho_0$, effectively control their radii. In this density regime, the PNM EoS can play an important role for testing the microscopic model Hamiltonians fit to $NN$ scattering data and few-body observables against astrophysical constraints. On the other hand, microscopic calculations of the EoS with reliable error estimates up to $2\rho_0$ provide useful insights on how measurements of the tidal polarizabilities from binary neutron-star mergers can unravel properties of matter at supra-nuclear densities~\cite{Tews:2018chv}.

In addition to the uncertainties arising from modeling the nuclear Hamiltonian, which can in principle be assessed by testing the order-by-order convergence of the chiral expansion~\cite{Epelbaum:2014efa}, microscopic calculations of the EoS are also affected by the approximations inherent to the method used for solving the many-body Schr\"odinger equation. To gauge them, we perform benchmark calculations of the energy per particle of pure neutron matter as a function of the baryon density using three independent many-body methods: the Brueckner--Bethe--Goldstone (BBG)~\cite{bbg1,bbg2}, the Fermi hypernetted chain/single-operator chain (FHNC/SOC)~\cite{FR75,PW79}, and the Auxiliary-field diffusion Monte Carlo (AFDMC)~\cite{Schmidt:1999lik}. In addition to the widely-used Argonne $v_{18}$ (AV18) NN potential~\cite{Wiringa:1994wb} -- and its simplified versions AV8$^\prime$ and AV6$^\prime$~\cite{wiringa02} -- we also consider the recently derived Norfolk NV2 $\chi$EFT NN forces~\cite{Piarulli:2014bda,Piarulli:2016vel}, which explicitly include the $\Delta$ isobar intermediate state. 

The scope of this work is not to achieve a realistic description of the EoS of PNM, which would require the inclusion of many-nucleon forces, as for instance in Refs.~\cite{Friedman:1981qw,WFF88,Akmal:1998cf,Lynn:2015jua,Bombaci:2018ksa}. We are rather mostly interested in quantitatively assessing the systematic error of the different many-body approaches and how this error depends upon the nuclear interaction of choice. The authors of Ref.~\cite{Baldo:2012nh} argued that the discrepancies among the methods are particularly susceptible to the spin-orbit components of the NN force. To identify and reduce these differences, we implement two major advancements in the AFDMC algorithm, both in the sampling procedure and in the way the fermion sign problem is controlled, in a similar fashion as recently done for atomic nuclei~\cite{Gandolfi:2014ewa,Lonardoni:2018nob}. The FHNC/SOC approach is also made more accurate by including classes of elementary diagrams that have been disregarded in earlier applications of the method. 

Recently, the scale dependence of both AV18 and the local $\chi$EFT interactions of Refs.~\cite{Gezerlis:2013ipa,Gezerlis:2014zia} has been investigated analyzing their predictions for NN scattering data and deuteron properties~\cite{Benhar:2019rro}. The main conclusion of that work is that phenomenological potentials appear to be best suited to study the high-density region of the EoS. Here, we extend this analysis comparing the energy per particle of PNM as obtained from both the Argonne and Norfolk NN interactions, relating their predictive power in describing the EoS at $\rho>\rho_0$ to their capability of reproducing NN scattering data as a function of the laboratory energy. 

The plan of this paper is as follows. The Argonne and Norfolk Hamiltonians are described in Sec.~\ref{sec:nuc_int}, where we also show the phase shifts predicted by the various NN potentials. The many-body methods employed for calculating the EoS of PNM are reviewed in Sec.~\ref{sec:mbm}, along with a detailed discussions of their technical improvements. The results obtained within the BBG, FHNC/SOC, and AFDMC approaches for the different Hamiltonians are benchmarked in Sec.~\ref{sec:results}. Finally, in Sec.~\ref{sec:conclusions} we summarize our findings and draw our conclusions.

\section{Nuclear interactions}
\label{sec:nuc_int}
In recent years local, configuration-space chiral interactions, well suited for use in QMC calculations of light-nuclei spectra and neutron-matter properties, have been derived by two groups~\cite{Gezerlis:2013ipa,Gezerlis:2014zia,Lynn:2014zia,Lynn:2015jua,Tews:2015ufa,Lynn:2017fxg,Piarulli:2017dwd}. In this paper, we will base our calculations on high-quality local potentials derived from a $\chi$EFT that explicitly includes---in addition to 
nucleons and virtual pions---virtual $\Delta$'s as degrees
of freedom~\cite{Piarulli:2014bda,Piarulli:2016vel,Piarulli:2017dwd,Baroni:2018fdn}.
The two-nucleon part ($N\!N$) of such local interactions is written as the sum of an electromagnetic interaction component, $v_{12}^{\rm EM}$, (as in Ref.~\cite{Wiringa:1994wb}), and a strong-interaction component, $v_{12}$, characterized by long- and short-range parts~\cite{Piarulli:2016vel}, respectively $v_{12}^{\rm L}$ and $v_{12}^{\rm S}$. The $v_{12}^{\rm L}$ part includes one-pion-exchange (OPE) and two-pion-exchange (TPE) terms up to next-to-next-to-leading order (N2LO) in the chiral expansion~\cite{Piarulli:2014bda}, derived in the static limit from leading and sub-leading $\pi\!N$ and $\pi\!N\!\Delta$ chiral Lagrangians. The $v_{12}^{\rm S}$ part, however, is described by contact terms up to next-to-next-to-next-to-leading order (N3LO)~\cite{Piarulli:2016vel}, characterized by 26 low-energy constants (LECs).
These interactions have been recently constrained to a large set  of $N\!N$-scattering data, as assembled by the Granada group~\cite{Perez:2013jpa}, including the deuteron ground-state energy and two-neutron scattering length. Particularly, we constructed two classes of $N\!N$ interactions, which only differ in the range of laboratory energy over which the fits were carried out, either 0--125 MeV in class I or 0--200 MeV in class II.
For each class, three different sets of cutoff radii $(R_{\rm S},R_{\rm L})$ were considered $(R_{\rm S},R_{\rm L})\,$=$\,(0.8,1.2)$ fm in set a, (0.7,1.0) fm in set b,
and (0.6,0.8) fm in set c, where $R_{\rm S}$ and $R_{\rm L}$ enter respectively the configuration-space cutoffs for the short- and long-range parts of the two-nucleon interaction~\cite{Piarulli:2016vel}. We are referring to these high-quality $N\!N$ interactions generically as the Norfolk (NV2) potentials, and denote those in class I as NV2-Ia, NV2-Ib, and NV2-Ic, and those in class II as NV2-IIa, NV2-IIb, and NV2-IIc.

The NV2 models were found to provide insufficient attraction, in Green's function Monte Carlo (GFMC) calculations, for the binding energies of light nuclei~\cite{Piarulli:2016vel}, thus confirming the insight realized in the early 2000's within the older (and less fundamental) meson-exchange phenomenology. To help remedy this shortcoming, we also constructed the leading three-nucleon ($3N$) interaction in $\chi$EFT, including $\Delta$ intermediate states. It consists~\cite{vanKolck:1994yi,Epelbaum:2002vt} of a long-range piece mediated by TPE and a short-range piece parametrized in terms of two contact interactions. The two $3N$ LECs, namely $c_D$ and $c_E$, have been obtained either by fitting exclusively strong-interaction observables~\cite{Lynn:2015jua,Tews:2015ufa,Lynn:2017fxg,Piarulli:2017dwd} or by relying on a combination of strong- and weak-interaction ones~\cite{Gazit:2008ma,Marcucci:2011jm,Baroni:2018fdn}. This last strategy is made possible by the relation, established in $\chi$EFT~\cite{Gardestig:2006hj}, between $c_D$ in the $3N$ interaction and the LEC in the $N\!N$ contact axial current~\cite{Gazit:2008ma,Marcucci:2011jm,Schiavilla:2017}, which allows one to use nuclear properties governed by either the strong or weak interactions to constrain simultaneously the $3N$ interaction and $N\!N$ axial current.  We designate the combinded $N\!N$ and $3N$ Norfolk potentials as the NV2+3 models.  

For the purpose of this paper, we will focus our attention on calculations of the EoS of neutron matter involving the NV2 local chiral $N\!N$ interactions and leave the NV2+3 models to future study.
Comparison will be made with the phenomenological AV18 potential~\cite{Wiringa:1994wb}.
Both the Argonne and Norfolk interactions are defined in coordinate space as
\begin{equation}
v_{ij} = \sum_{p=1}^{N} v^p(r_{ij})O^p_{ij}
\label{eq:vNN}
\end{equation}
with $r_{ij}= |{\bf r}_i - {\bf r}_j|$.  For the Argonne potential $N=18$
-- hence the name Argonne $v_{18}$ or AV18 -- while the NV2 potentials 
have $N=16$.  
The bulk of the NN interaction is encoded in the first eight operators
\begin{equation}
O^{p=1-8}_{ij}= [1, \sigma_{ij},S_{ij},\mathbf{L}\cdot\mathbf{S}]\otimes [1,\tau_{ij} ] \, ,
\label{eq:oper}
\end{equation}%
which are the same for both AV18 and the NV2s.
In the above equation we introduced $\sigma_{ij}={\bm \sigma_i}\cdot{\bm \sigma_j}$ and $\tau_{ij}={\bm \tau_i}\cdot{\bm \tau_j}$ with ${\bm \sigma}_i$ and ${\bm \tau}_i$ being the Pauli matrices acting in the spin and isospin space. The tensor operator is given by
\begin{equation}
S_{ij}= \frac{3}{r^2_{ij}}({\bm \sigma}_i\cdot {\bf r}_{ij})({\bm \sigma}_j \cdot{\bf r}_{ij})- \sigma_{ij}\, ,
\end{equation}
while the spin-orbit contribution is expressed in terms of the relative angular momentum $\mathbf{L}=\frac{1}{2 i} (\mathbf{r}_i -\mathbf{r}_j) \times (\nabla_i - \nabla_j)$ and the total spin $\mathbf{S}=\frac{1}{2}({\bm \sigma}_j+{\bm \sigma}_j)$ of the pair. 
For AV18 there are six additional charge-independent operators corresponding to $p=9-14$ that are quadratic in $\mathbf{L}$, while the $p=15-18$ are charge-independence breaking terms.
In contrast, the NV2 potentials have three charge-independent operators quadratic in $\mathbf{L}$, and five charge-independence breaking terms.

It is useful to define simpler versions of the AV18 and NV2 potentials with 
fewer operators: a $v^\prime_8$ with the eight operators of Eq.~\eqref{eq:oper}
and a $v^\prime_6$ without the $\mathbf{L}\cdot\mathbf{S}\otimes [1,\tau_{ij} ]$
terms~\cite{pudliner97,wiringa02}.  
The $v^\prime_8$ is a reprojection (rather than a simple truncation)
of the strong-interaction potential that reproduces the charge-independent
average of $^1$S$_0$, $^3$S$_1$-$^3$D$_1$, $^1$P$_1$, $^3$P$_0$, $^3$P$_1$, 
and (almost) $^3$P$_2$ phase shifts by construction, while overbinding the
deuteron by 18 keV due to the omission of electromagnetic terms.
The $v^\prime_6$ is (mostly) a truncation of $v^\prime_8$ which reproduces
$^1$S$_0$ and $^1$P$_1$ partial waves, makes a slight adjustment to (almost)
match the $v^\prime_8$ deuteron and $^3$S$_1$-$^3$D$_1$ partial waves,
but will no longer split the $^3$P$_J$ partial waves properly.
We will refer to these variations of the Argonne potential as AV8$^\prime$ 
and AV6$^\prime$.

In strongly degenerate systems of fermions, such as the low-temperature nucleonic matter forming the interior of neutron stars, collisions primarily involve nucleons occupying states close to the Fermi surface. As a consequence, in the case of head-on scattering, a relation can be easily established between the kinetic energy of the beam particle in the lab frame, $E_{\rm lab}$, and the Fermi energy $E_F$, which in turn is simply related to the baryon density $\rho$. The resulting expression in PNM is 
\begin{equation}
E_{\rm lab} = 2 E_{\rm cm} = 4 E_F = \frac{2\hbar^2}{m} (3\pi^2\rho)^{2/3}
\label{eq:Ecm}
\end{equation}

\begin{figure}[h!]
\centering
\includegraphics[width=\columnwidth]{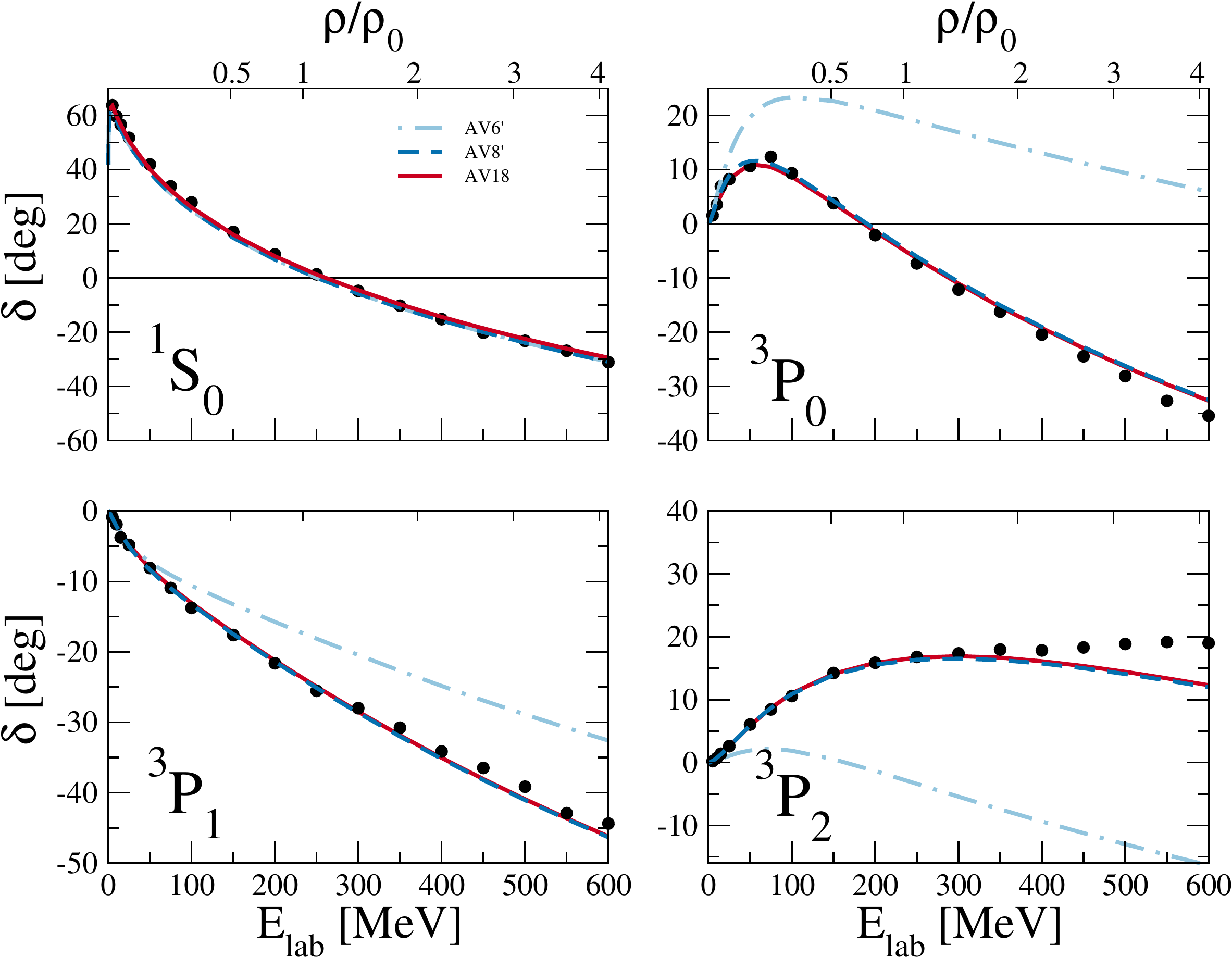}
\caption{Neutron-proton scattering phase shifts in the $^1$S$_0$, $^3$P$_0$, $^3$P$_1$, and $^3$P$_2$ channels, as a function of kinetic energy of the beam particle in the laboratory frame (bottom axis). The corresponding densities of PNM -- in units of $\rho_0$ -- are given in the top axis. The long-dashed-dotted, the dash- an the solid lines represent the  AV6$^\prime$, AV8$^\prime$, and AV18 predictions, while the solid dots are from the SM16 solution of Ref.~\cite{Workman:2016ysf}.}
\label{fig:ps_argonne}
\end{figure}

\begin{figure}[h!]
\centering
\includegraphics[width=\columnwidth]{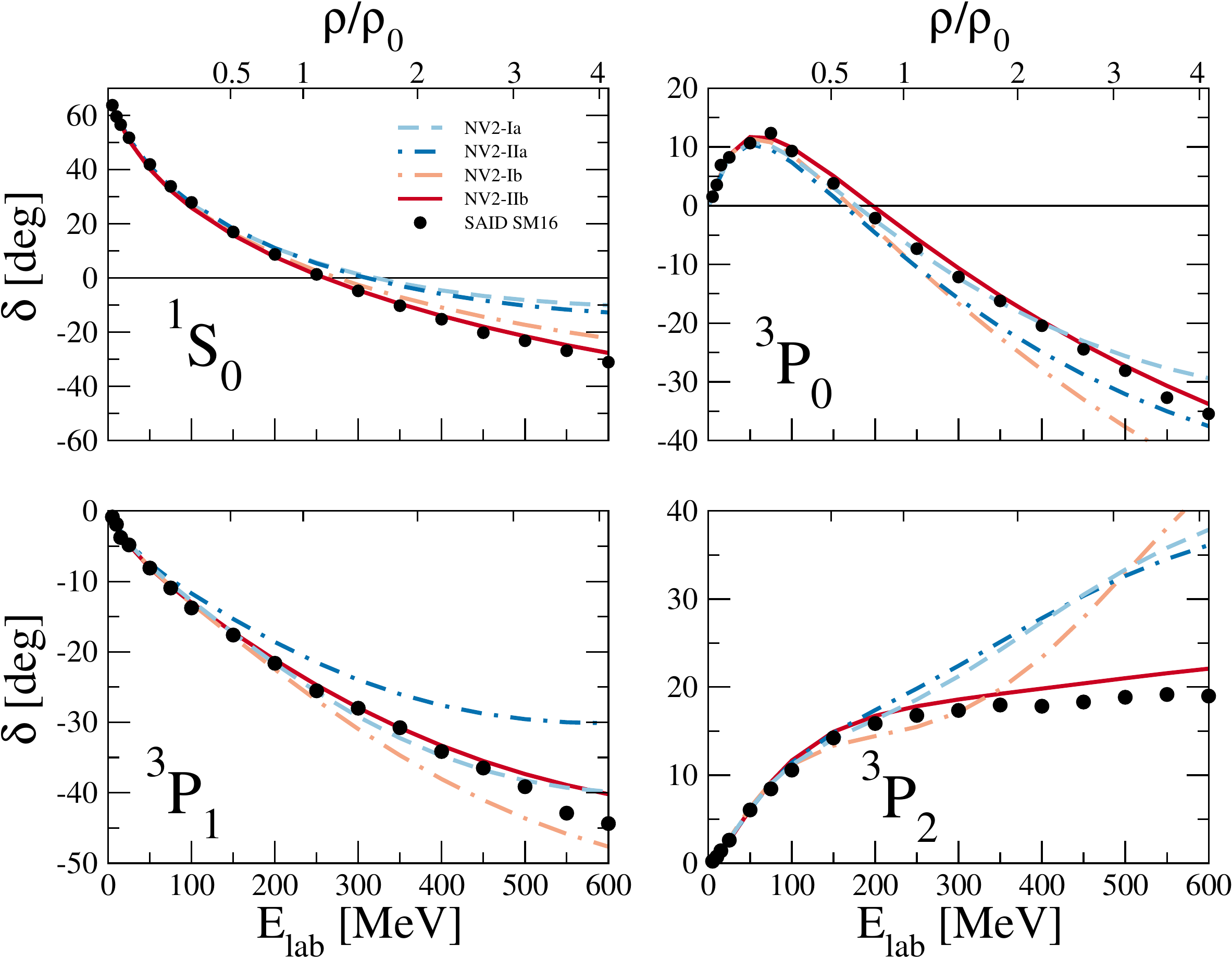}
\caption{Same as Fig.~\ref{fig:ps_argonne}, but for the NV2 $\Delta$-full local $\chi$EFT interactions. The dash- and short-dashed-dotted lines represent the predictions for the NV2-Ia and NV2-IIa models, having $R_S=0.8$ fm and, respectively, constrained to $N\!N$-scattering data up to 125 and 200 MeV laboratory energy. The long-dashed-dotted and the solid lines show results for the NV2-Ib and NV2-IIb models, having $R_S=0.7$ fm and, respectively, constrained to $N\!N$-scattering data up to 125 and 200 MeV laboratory energy. The solid dots are from the SM16 solution of Ref.~\cite{Workman:2016ysf}.}
\label{fig:ps_chiral}
\end{figure}

In Ref.~\cite{Benhar:2019rro} the above expression has been utilized to gauge the predictive power of NN potential models in describing the high-density regime of PNM. Along the same line, Fig.~\ref{fig:ps_argonne} illustrates the energy dependence of the proton-neutron scattering phase shifts in the $^1$S$_0$, $^3$P$_0$, $^3$P$_1$, and $^3$P$_2$ partial waves comparing the AV6$^\prime$, AV8$^\prime$, and AV18 potentials with the analysis of Ref.~\cite{Workman:2016ysf}. In Fig.~\ref{fig:ps_chiral} we show the predictions for the same quantities obtained from the set of NV2 $\Delta$-full local $\chi$EFT interactions discussed above. The density of PNM obtained from Eq.~\eqref{eq:Ecm} with $E_{\rm lab} = 2E_{\rm cm}$ is reported on the top axis of the figures in units of the nuclear saturation density $\rho_0=0.16$ fm$^{-3}$. The AV18 interaction provides an accurate description of the scattering data up to $E_{\rm lab} \simeq 600$ MeV and appears to be applicable to describe properties of PNM at least up to $\rho \simeq 4\rho_0$. 

As opposed to the AV18 potential, $\chi$EFT models, which  are based on low momentum expansion, are intrinsically limited in describing dense systems, in which $NN$ interactions involve high energies.
In addition, chiral potentials depend on regulator function that smoothly cuts off one- and two-pion exchange interactions at short distances, and the choice of such short range cutoffs, here defined as $R_S$, plays a crucial role in describing the short-range dynamics in dense systems. In Fig.~\ref{fig:ps_chiral}, we notice that the chiral model NV2-IIb, which is fitted at higher energies ($E_{\rm lab}$ = 200 MeV) with a hard cutoff ($\sim$ 600 MeV in momentum-space) achieves a better description of the $S$- and $P$- phase shifts up to $\sim$ 600 MeV. Such model performs very closely to the AV18, which has been fitted up to the pion  production threshold ($E_{\rm lab}$ = 350 MeV) with a very hard core cutoff ($\sim$ 1 GeV).

\section{Many-body methods}
\label{sec:mbm}
\subsection{Brueckner--Bethe--Goldstone many body theory}
The Brueckner--Bethe--Goldstone (BBG) many-body theory (see eg. \cite{bbg1,bbg2}) is based on a 
linked cluster expansion (the so-called hole-line expansion) of the energy per nucleon $E/A$ of nuclear matter.   
The various terms of the expansion can be represented by Goldstone diagrams \cite{goldstone57} 
grouped according to the number of independent hole-lines ({\it i.e.} lines representing empty single 
particle states in the Fermi sea). 
The basic ingredient in this approach is the Brueckner reaction matrix $G$ \cite{brueck1,brueck2} 
which sums, in a closed form the infinite series of the so-called ladder-diagrams and allows to deal with 
the short-range strongly repulsive part of the nucleon-nucleon interaction.   
The G-matrix can be obtained by solving the Bethe--Goldstone equation \cite{BG57}  
\begin{equation}
 G(\omega) =  V  + V  \sum_{k_a,k_b} 
\frac{\mid{\bf k_a},{\bf k_b}\rangle \, Q\, \langle{\bf k_a},{\bf k_b}\mid}
     {\omega - \epsilon(k_a) - \epsilon(k_b) + i\eta }\, G(\omega) \;,
\label{bg} 
\end{equation}
where $V$ is the bare NN interaction, the quantity $\omega$ is the so called starting energy. 
In the present work we consider spin unpolarized neutron matter, thus in equation (\ref{bg}) and in the 
following equations we drop the spin indices to simplify the mathematical notation.    
The Pauli operator $\mid{\bf k_a},{\bf k_b}\rangle Q \langle {\bf k_a},{\bf k_b}\mid$ 
projects on intermediate scattering states in which the momenta ${\bf k_a}$ and ${\bf k_b}$
of the two interacting neutrons are above their Fermi momentum $k_{F}$ since single particle states 
with momenta smaller that this value are occupied by the neutrons of the nuclear medium.   
Thus the Bethe--Goldstone equation describes the scattering of two nucleons (two neutrons in our case) 
in the presence of other nucleons, and the Brueckner $G$-matrix represents the effective interaction between 
two nucleons in the nuclear medium and properly takes into account the short-range correlations arising from 
the strongly repulsive core in the bare NN interaction.

The single-particle energy $\epsilon(k)$ of a neutron with momentum ${\bf k}$, appearing in the energy 
denominator of the Bethe--Goldstone equation (\ref{bg}), is given by 
\begin{equation}
       \epsilon(k) = \frac{\hbar^2 k^2}{2m} + U(k) \ ,
\label{spe}
\end{equation}
where $U(k)$ is a single-particle potential which represents the mean field felt by a neutron  
due to its interaction with the other neutrons of the medium. 
In the Brueckner--Hartree--Fock (BHF) approximation of the BBG theory, $U(k)$ 
is calculated through the real part of the $G$-matrix \cite{BBP63,HM72} and is given by 
\begin{equation}
 U(k) = \sum_{k^\prime \leq k_{F}} 
            \mbox{Re} \ \langle {\bf k},{\bf k^\prime} \mid 
               G(\omega^*) \mid {\bf k},{\bf k^\prime} \rangle_A  \;,
\label{spp}
\end{equation}
where the sum runs over all neutron occupied states, the starting energy is 
$\omega = \omega^* \equiv \epsilon(k) + \epsilon(k')$ (i.e. the G-matrix is calculated on-the-energy-shell) 
and the matrix elements are properly antisymmetrized. 
We make use of the so-called continuous choice \cite{jeuk+67,gra87,baldo+90,baldo+91} for 
the single-particle potential $U(k)$  when solving the Bethe--Goldstone equation.    
As it has been shown in Ref. \cite{song98,baldo00}, the contribution of the three-hole-line diagrams 
to the energy per nucleon $E/A$ is minimized in this prescription for the single particle potential and 
a faster convergence of the hole-line expansion for $E/A$ is achieved with respect to the so-called 
gap choice for $U(k)$.  

In this scheme Eqs.\ (\ref{bg})--(\ref{spp}) have to be solved self-consistently 
using an iterative numerical procedure. 

Once a self-consistent solution is achieved, the energy per nucleon of the system 
can be evaluated in the BHF approximation of the BBG hole line-expansion 
and it is given by    
\begin{equation}
 \frac{E}{A} = \frac{1}{A} \sum_{k < k_{F}}
    \left(\frac{\hbar^2 k^2}{2m} + \frac{1}{2} U(k) \right) \ .
\label{bea}
\end{equation}

In this approach the two-body interaction ${V}$ is the only physical input for 
the numerical solution Bethe--Goldstone equation.  

Making the usual angular average of the Pauli operator and of the energy denominator \cite{gra87,baldo+91}, 
the Bethe--Goldstone equation (\ref{bg}) can be expanded in partial waves. 
In all the calculations performed in this work, we have considered partial wave contributions up to a total 
two-body angular momentum $J_{max} = 11$.  
We have verified that the inclusion of partial waves with $J_{max} > 11$ does not appreciably 
change our results.

\subsection{Fermi hypernetted chain / single-operator chain method}
In absence of interactions, a uniform system of $A$ non-interacting neutrons can be described as a Fermi gas at zero temperature, and its ground state wave function reduces to the Slater determinant of orbitals associated with the single-particle states belonging to the Fermi sea
\begin{equation}
\Phi(X)=\mathcal{A}[\,\phi_{n_1}(x_1)\dots\phi_{n_A}(x_A)\,]\, .
\label{def:phi}
\end{equation}
In the above equation $X=\{x_1,\dots,x_A\}$, where the generalized coordinate $x_i\equiv\{\mathbf{r}_i,s_i\}$ represents both the position $R=\mathbf{r}_1,\dots,\mathbf{r}_A$ and the spin $S=s_1,\dots,s_A$, variables of the $i$-th nucleon while $n_i$ denotes the set of quantum numbers specifying the single particle state. Translational invariance imposes that the single-particle wave functions be plane waves
\begin{equation}
\phi_{n_i}(x_i)= \frac{1}{\sqrt{\Omega}} {\rm e}^{i\mathbf{k_i}\mathbf{r}_i} \chi_{\sigma_i}(s_i)
\end{equation}
In the above equations, $\Omega$ is the normalization volume, $\chi_{\sigma_i}(s_i)$ is the spinor of the neutron and $|\mathbf{k}_i|<k_F=(3\pi^2\rho)^{1/3}$. Here $k_F$ is the Fermi momentum and $\rho$ the density of the system. 

The variational ansatz of the Fermi hypernetted chain (FHNC) and single-operator chain (SOC) formalism emerges as a generalization of the Jastrow theory of Fermi liquids~\cite{jastrow55,PW79}
\begin{equation}
|\Psi_T\rangle=\frac{\hat{F}|\Phi\rangle}{\langle\Phi|\hat{F}^\dagger \hat{F}|\Phi\rangle^{1/2}}\, .
\label{eq:psi_FHNC}
\end{equation}
where $|\Phi\rangle$ is the Slater determinant of Eq.~\eqref{def:phi} and 
\begin{equation}
\mathcal{\hat{F}}(x_1,\dots,x_A)=\mathcal{S} \left( \prod_{j>i=1}^A \hat{F}_{ij} \right)
\label{eq:Foperator}
\end{equation}
is the correlation operator. The spin-isospin structure of $F_{ij}$ reflects that of the nucleon-nucleon potential of Eq.~\eqref{eq:vNN}
\begin{equation}
\hat{F}_{ij} = \sum_{p=1}^8 f^{p}(r_{ij})\hat{O}^{p}_{ij} \ .
\label{eq:Foperator_v8prime}
\end{equation}
Since, in general, $[\hat{O}^{p}_{ij},\hat{O}^{q}_{ik}] \neq 0$, the symmetrization operator $\mathcal{S}$ is needed to fulfill the requirement of antisymmetrization of the wave-function.
The $f^{p}(r_{ij})$ are finite-ranged functions, with the conditions
\begin{align}
f^{p}(r\geq d_p) &= \delta_{p1}\, , \nonumber \\
\frac{df^p(r)}{dr}\Big|_{r=d_p} &= 0\, .
\label{eq:heal}
\end{align}
where the $d_p$ are ``healing distances''.  Consequently, 
the correlation operator of Eq.~\eqref{eq:Foperator} respects the cluster property: if the system is split in two (or more) subsets of particles that are moved far away from each other, the $\hat{F}$ factorizes into a product of two factors in such a way that only particles belonging to the same subset are correlated. For instance, consider two subsets, say $i_1, . . . i_M$ and $i_{M+1}, . . . i_A$. The cluster property implies
\begin{equation}
\mathcal{\hat{F}}(x_1,\dots,x_A)=\mathcal{\hat{F}}(x_{i_1},\dots,x_{i_M}) \mathcal{\hat{F}}(x_{i_{M+1}},\dots,x_A)
\label{eq:cluster}
\end{equation}

The radial functions $f^{p}(r_{ij})$ are determined by minimizing the energy expectation value
\begin{equation} 
 E_V=\langle\Psi_T|H|\Psi_T\rangle \geq E_0\, , 
\label{eq:ev}
\end{equation}
which provides an upper bound to the true ground state energy $E_0$. The cluster property allows one to expand the expectation value of the Hamiltonian -- and of other many-body operators -- between correlated states in a sum of cluster contributions involving an increasing number of particles. 

The energy expectation value in matter is evaluated using a diagrammatic 
cluster expansion and a set of 29 coupled integral equations, which effectively
make partial summations to infinite order -- the FHNC/SOC approximation~\cite{PW79}.  
This is a generalization of the original hypernetted chain (HNC) method for 
Bose systems developed by van Leeuwen, Groeneveld, and de Boer~\cite{vLGdB59},
which requires the solution of a single integral equation, and the 
corresponding extension for spin-isospin independent Fermi systems by Fantoni
and Rosati~\cite{FR75}, which requires four coupled integral equations.
The integral equations are used to generate two- and three-body distribution
functions $g_2(r_{ij}) \equiv g_{ij}$ and 
$g_3({\bf r}_{ij},{\bf r}_{ik}) \equiv g_{ijk}$, 
which can then be used to evaluate the energy or other operators.  

For the pure Jastrow case, we evaluate the Pandharipande-Bethe~\cite{PB73}
expression for the energy:
\begin{equation}
E_{PB} = T_F + W + W_F + U + U_F \ ,
\label{eq:epb}
\end{equation}
where $T_F$ is the Fermi gas kinetic energy. The only terms for a Bose system 
are
\begin{align}
  W &= \frac{\rho}{2} \int \Big( v_{ij} - \frac{\hbar^2}{m}
        \frac{\nabla^2 F_{ij}}{F_{ij}} \Big) g_{ij} d^3r_{ij} \ , \nonumber \\
  U &= - \frac{\hbar^2}{2m} \frac{\rho^2}{4} \int
  \Big(\frac{{\bf\nabla}_i F_{ij}\cdot {\bf\nabla}_i F_{ik}}{F_{ij}F_{ik}}\Big)
             g_{ijk} d^3r_{ij} d^3r_{ik}\, ,
\end{align}
while $W_F$, $U_F$ are additional 
two- and  three-body kinetic energy terms present due to the Slater determinant.
Alternately we use the Jackson-Feenberg~\cite{JF62} energy expression
\begin{eqnarray}
E_{JF} &=& T_F + W_B + W_\phi + U_\phi \ , \\
  W_B &=& \frac{\rho}{2} \int \Big[v_{ij}- \frac{\hbar^2}{2m}
\Big(\frac{\nabla^2 f_{ij}}{f_{ij}} - \frac{(\nabla_i f_{ij})^2}{f^2_{ij}} \Big) \Big] \ ,
\label{eq:ejf}
\end{eqnarray}
where $W_B$ is the boson term and $W_\phi$ and $U_\phi$ are kinetic energy
terms involving the Slater determinant.
In principle, these energies should be equivalent, but in practice there are
differences due to the FHNC/SOC approximation to the distribution functions.
We take the average $E_V = (E_{PB} + E_{JF})/2$ as our energy expectation value
and the difference $\delta E_V = |E_{PB} - E_{JF}|/2$ as an estimate of the 
error in the calculation.

The FHNC two-body distribution function can be written as:
\begin{eqnarray}
g_{ij} &=& f^2 \Big[ (1+G_{de}+{\mathcal E}_{de})^2 + G_{ee}+{\mathcal E}_{ee} \nonumber \\
       &-& \nu(G_{cc}+{\mathcal E}_{cc}-\ell/\nu)^{2} \Big]
           \exp(G_{dd}+{\mathcal E}_{dd}) \ .
\label{eq:gij}
\end{eqnarray}
where the chain functions $G_{xy}$ are sums of nodal diagrams, with direct 
($d$), exchange ($e$) or circular exchange ($c$) end points and
${\mathcal E}_{xy}$ are elementary diagrams.
An example of the structure of the integral equations is:
\begin{align}
G_{dd,ij} &= \rho\int d^{3}r_{k} \left[ ( X_{dd,ik} + X_{de,ik} ) S_{dd,kj} \right. \nonumber\\
&\left. + X_{dd,ij}S_{de,kj} \right] \ ,
\end{align}
where $S_{dd} = f^{2}\exp(G_{dd}+E_{dd}) - 1$ is a two-point superbond
and $X_{dd} = S_{dd} - G_{dd}$ is a link function.

The introduction of spin-isospin correlations with operators that do not
commute complicates the calculation.  Fortunately, the first six operators 
$p=1,6$ form a closed spin-isospin algebra, allowing single continuous chains 
of operator links -- the SOCs -- to be evaluated.  
These involve five chain functions $G^p_{xy}$ for each of the five
operators $p=2-6$, with $xy = dd, de, ee, ca, cb$ in addition to the four
Jastow chain functions in Eq.\eqref{eq:gij}, making the total of 29 coupled
integral equations to be solved.
There are significant contributions from unlinked diagrams in the SOC
cluster expansion, but these can be accomodated by means of ``vertex'' 
corrections, as discussed in Ref.~\cite{PW79}.
Additional higher-order corrections coming from (parallel) multiple operator 
chains and rings are also calculated, as discussed in Refs~\cite{WFF88}.
Spin-orbit correlations, i.e., $p=7,8$, cannot be ``chained'' so they are
treated explicitly only at the two- and three-body cluster level.

In standard FHNC calculations, the elementary diagrams of Eq.~\eqref{eq:gij} 
are generally neglected.  Inclusion of the leading four-body elementary
diagram leads to the FHNC/4 approximation~\cite{Zabo77}, while additional 
contributions have been studied in liquid atomic helium systems~\cite{UP82}.
In the present work we include many central ($p=1$) ${\mathcal E}_{xy}$ 
diagrams, beyond the FHNC/4 approximation, by introducing three-point
superbonds $S_{xyz}$, such as
\begin{eqnarray}
S_{ddd,123} &=& \rho\int d^{3}r_{4} \left\{ S_{dd,14}S_{dd,24}
                   ( S_{dd,34} + S_{de,34} ) \right. \nonumber \\
  &+& ( S_{dd,14}S_{de,24} + S_{de,14}S_{dd,24} ) S_{dd,34} \left.\right\} \ ,
\end{eqnarray}
and then evaluating
\begin{eqnarray}
   {\mathcal E}_{dd,12} &=& \frac{1}{2}\rho\int d^{3}r_{3}
                 \left\{ S_{ddd,132}\left[ S_{dd,13}( S_{dd,32} + S_{de,32} )
                 \right.\right. \nonumber\\
             &+& S_{de,13}S_{dd,32} \left.\right]
               + S_{ded,132}S_{dd,24}S_{dd,32} \left.\right\} \ .
\end{eqnarray}
With six $S_{xyz}$, where $xyz$ = $ddd$, $dde$, $dee$, $eee$, $ccd$, and $cce$,
many elementary diagrams at the four-, five-, and higher-body level contributing
to $g_{ij}$ and $g_{ijk}$ can be evaluated.  These central elementary diagrams 
also dress the SOCs.

In matter calculations, the correlations of Eq.\eqref{eq:heal} are generated by 
solving a set of coupled Euler-Lagrange equations in different pair-spin and
isospin channels for $S=0,1$ and $T=0,1$.  For pure neutron matter, only
$T=1$ channels are needed, leaving a single-channel equation for $S=0$,
producing a singlet correlation, and a triple-channel equation for $S=1$, which
produces triplet, tensor, and spin-orbit correlations.  The singlet and
triplet correlations are then projected into central and $\sigma_{ij}$
combinations.  Three (increasing) healing distances are used: 
$d_s$ for the singlet correlation, $d_p$ for the triplet and spin-orbit,
and $d_t$ for the tensor.

Additional variational parameters are the quenching factors $\alpha_p$ whose introduction simulates modifications of the two--body potentials entering in the Euler--Lagrange differential equations arising from the screening induced by the presence of the nuclear medium 
\begin{equation}
v_{ij}=\sum_{p=1}^8 \alpha_p v^{p}(r_{ij})O^{p}_{ij}\, ,
\end{equation}
whereas the full potential is used when computing the energy expectation value.
In practice we use just two such parameters: $\alpha_{p=1} = 1$ and 
$\alpha_{p=2,8} = \alpha$.  In addition, the resulting correlation functions 
$f^p$ may be rescaled according to  
\begin{equation}
F_{ij}=\sum_{p=1}^8 \beta_p f^{p}(r_{ij})O^{p}_{ij}\; ,
\end{equation}
with $\beta_{p=1} = 1$, $\beta_{p=2,4+7,8} = \beta_\sigma$, and 
$\beta_{p=5,6} = \beta_t$.  However, these are usually invoked only in the
presence of three-body forces.  For the present work, the variational
parameters are the three healing distances and one quenching factor.
These are varied at each density with a simplex search routine to minimize
the energy.  

One measure of the convergence of the FHNC/SOC integral equations is that the
volume integral of the correlation hole from the central part of the two-body 
distribution function $g_{ij}$ (which has operator components like the 
$\hat{F}_{ij}$ of Eq.~\eqref{eq:Foperator_v8prime}) should be unity.
To help guarantee that the variational parameters entering the FHNC/SOC
correlations are well behaved, we minimize the energy plus a constant times 
the deviation of the volume integral from unity:
\begin{equation}
 E + C \left\{ 1 + \rho\int d^{3}r [ g^c(r)-1 ] \right\}^2 \ , \nonumber
\end{equation}
as discussed in Ref.\cite{WFF88}.  A value of $C=1000$ MeV is sufficient
to limit the violation of this sum rule to 1\% or less at normal density,
and 3\% or less at twice normal density for all the potentials considered
here.  There is a related sum rule for the isospin component $g^\tau$ 
that applies in symmetric nuclear matter, but there is no sum rule for the 
spin correlation hole for realistic potentials with tensor forces.

\subsection{Auxiliary-field diffusion Monte Carlo}
Over the last two decades, the auxiliary-field diffusion Monte Carlo (AFDMC) method~\cite{Schmidt:1999lik} has become a mainstay for neutron-matter calculations~\cite{Tews:2015ufa,Lynn:2015jua,Tews:2018kmu}. Within the AFDMC, properties of the infinite uniform system are simulated with a finite number of neutrons obeying periodic-box boundary condition (PBC). The trial wave function is a simplified version of the one reported in Eq.~\eqref{eq:psi_FHNC}
\begin{align}
\Psi_T(X)=\langle X|\Psi_T\rangle=\langle X|\Bigg(\prod_{i<j} f^c(r_{ij})\Bigg)|\Phi\rangle .
\label{eq:psi_T}
\end{align}
The anti-symmetric mean-field part $|\Phi\rangle$ is the Slater determinant of Eq.~\eqref{def:phi}. In order to satisfy the PBC, the single-particle wave vector is discretized as
\begin{equation}
\mathbf{k}_i=\frac{2\pi}{L} \{ n_x,n_y,n_z\}\quad,\quad n_i=0,\pm 1,\pm 2,\dots\, ,
\end{equation}
$L$ being the size of the simulation box. When not otherwise specified, in our simulations we typically employ $A=66$ neutrons in a box. Finite-size errors in PNM simulations have been investigated in Ref.~\cite{Fantoni:2008jd,Gandolfi:2009fj} by comparing the twist averaged boundary conditions with the PBC. Remarkably, the PBC energies of 66 neutrons differ by no more than 2\% from the asymptotic value calculated with twist averaged boundary conditions. This essentially follows from the fact that the kinetic energy of 66 fermions approaches the thermodynamic limit very well. Additional finite-size effects due to the tail corrections of two- and three-body potentials are accounted for by summing the contributions given by neighboring cells to the simulation box~\cite{Sarsa:2003zu}.

The spin-independent correlation ansatz of Eq.~\eqref{eq:psi_T} has proven to be inadequate to treat atomic nuclei and infinite nucleonic matter comprised of both neutrons and protons. In fact, the expectation value of the tensor components of the NN potential, which is large for neutron-proton pairs in the $T=0$ channel, is nearly zero when tensor correlations are not included in $\Psi_T(X)$. To overcome these difficulties, a linearized version of spin-dependent two-body correlations, in which only one pair of nucleons is correlated at a time, was first implemented in the AFDMC method in Ref.~\cite{Gandolfi:2014ewa}. Very recently, the trial wave function has been further improved by including quadratic pair correlations~\cite{Lonardoni:2018nob}. These more sophisticated wave functions have enabled a number of remarkably accurate AFDMC calculations, in which properties of atomic nuclei with up to $A=16$ nucleons~\cite{Lonardoni:2017hgs} have been investigated utilizing the local $\chi$EFT interactions of Ref.~\cite{Gezerlis:2013ipa,Lynn:2015jua}.

Analogously to the FHNC case, the two-body correlation functions are obtained by minimizing the two-body cluster contributions of the energy per particle, solving the same set of coupled Euler-Lagrange equations. However, since in $\Psi_T$ we only retain spin-independent terms, we found that replacing $f^c(r_{ij}) \to f^c(r_{ij})+\beta_\sigma f^\sigma(r_{ij})$,  $\beta_\sigma$ being a variational parameter, provides a better variational energy than when $\beta_\sigma=0$. The relatively simple trial wave function of Eq.~\eqref{eq:psi_T} is completely determined by three variational parameters: $\beta_\sigma$, the spin-isospin potential quencher $\alpha_{p=2,8} = \alpha$, and the central healing distance $d_c$, as for simplicity we assume $d_s=d_p=d_c$. As a consequence, it is unnecessary to use advanced optimizations algorithms, such as the ``stochastic reconfiguration''~\cite{Sorella:2005} or the ``linear method''~\cite{Toulouse:2007} algorithm, to minimize the variational energy. 

AFDMC is an extension of standard Diffusion Monte Carlo algorithms, in which the ground-state $\Psi_0$ of a given Hamiltonian is projected out from the starting trial wave function using an imaginary-time evolution
\begin{equation}
| \Psi_0\rangle = \lim_{\tau \to \infty} e^{-(H-E_T)\tau} | \Psi_T\rangle
\end{equation}
In the above equation $\tau$ is the imaginary time, and $E_T$ is a parameter used to control the normalization.  For strongly interacting systems, the direct computation of the propagator $e^{-(H-E_0)\tau}$ involves prohibitive difficulties. For small imaginary times $\delta \tau = \tau/N$, with $N$ being a large number, one can compute the short-time propagator, and the full propagation can be recovered inserting complete sets of states. The propagated wave function then reads
\begin{align}
\langle X_N | \Psi(\tau)\rangle &= \prod_{i=1}^{N-1} \int dX_i \langle X_N | e^{-(H-E_0)\delta\tau} |X_{N-1}\rangle \dots \nonumber\\
& \times \langle X_2 | e^{-(H-E_0)\delta\tau}|  X_1\rangle \langle X_1| \Psi_T\rangle\, .
\end{align}
By using the  Suzuki-Trotter decomposition to order $\delta\tau^3$, the short-time propagator can be cast in the form
\begin{align}
G(X,X^\prime,\delta\tau) &=  \langle X| e^{-(H-E_0)\delta\tau}|  X^\prime \rangle \nonumber\\
&\simeq \langle X | e^{-V \frac{\delta\tau}{2}} e^{-T \delta\tau} e^{-V \frac{\delta\tau}{2}}  | X^\prime \rangle\,.
\label{eq:prop}
\end{align}
In the above equation, $V$ is the nuclear potential and $T$ is the nonrelativistic kinetic energy, giving rise to the free propagator
\begin{align}
G_0(X,X^\prime,\delta\tau)&= \langle X| e^{-T \delta\tau}|X^\prime \rangle \nonumber\\
&=\left(\frac{m}{2 \pi\delta\tau}\right)^\frac{3A}{2} e^{-\frac{m(R-R^\prime)^2}{2\delta\tau}} \delta(S-S^\prime)\,,
\label{eq:T_prop}
\end{align}
Monte Carlo techniques are used to sample the paths $X_i$. In practice, a set of configurations, typically called {\it walkers,} are simultaneously evolved in imaginary time, and then used to calculate observables once convergence is reached. 

Within the Green's function Monte Carlo (GFMC) method used in light nuclei, the positions of the particles are sampled, but the full sum over the spin-isospin degrees of of freedom is retained, leading to an exponential growth of the computational cost with $A$.  The AFDMC method overcomes this limitation using a spin-isospin basis given by the outer product of single-nucleon spinors
\begin{equation}
|S\rangle = |s_1\rangle \otimes |s_2\rangle \dots \otimes |s_A\rangle \, . 
\end{equation}
Realistic nuclear potentials, such the ones employed in this work, contain quadratic spin/isospin operators. In order to preserve the single-particle representation, the short-time propagator is linearized utilizing the Hubbard-Stratonovich transformation
\begin{equation}
e^{-\lambda O^2 \delta\tau/2} = \frac{1}{\sqrt{2\pi}} \int_{-\infty}^\infty dx e^{-x^2/2} e^{x\sqrt{- \lambda \delta\tau}\, O}
\label{eq:V_prop}
\end{equation}
where $x$ are the {\it auxiliary fields} and the operators $O$ are obtained as follows. The first six terms defining the NN potential of Eq.~\eqref{eq:vNN} can be conveniently separated in a spin-isospin dependent $V_{SD}$ and spin-isospin independent $V_{SI}$ contributions.  Since in purely neutron systems $\tau_{ij}=1$, $V_{SD}$ can be cast in the form 
\begin{equation}
V_{SD}=\frac{1}{2} \sum_{i\alpha j \beta} A_{i\alpha, j\beta} \sigma_i^\alpha \sigma_j^\beta = \frac{1}{2}\sum_{n=1}^{3A} O^2_{n} \lambda_n\, ,
\end{equation}
where the operators $O_{n}$ are defined as
\begin{equation}
O_{n}=\sum_{i,\alpha} \sigma_i^\alpha \psi^n_{i\alpha}
\end{equation}
In the above equations $\lambda_n$ and  $\psi^n_{i\alpha}$ are the eigenvalues and eigenvectors of the matrix $A$. The spin-orbit term of the NN potentials is implemented in the propagator as described in Refs.~\cite{Sarsa:2003zu} and  appropriate counter terms are included to remove the spurious contributions of order $\delta\tau$.

Importance sampling techniques are routinely implemented in the AFDMC -- in both the spatial coordinates and spin-isospin configurations -- to drastically improve the efficiency of the algorithm. To this aim, the propagator of Eq.~\eqref{eq:prop} is modified as
\begin{align}
G_I(X,X^\prime) = G_I(X,X^\prime) \frac{\Psi_I(X^\prime)}{\Psi_I(X)}\,.
\end{align}
At each time-step, each walker is propagated sampling a $3A$-dimensional vector to shift the spatial coordinates and a set of auxiliary fields $\mathcal{X}$ from Gaussian distributions. To remove the linear terms coming from the exponential of Eqs.~\eqref{eq:T_prop},~\eqref{eq:V_prop}, in analogy to the GFMC method, we consider four weights, corresponding to separately flipping the sign of the spatial moves and spin-isospin rotations
\begin{equation}
w_i=\frac{\Psi_I(\pm R^\prime, S^\prime(\pm \mathcal{X}))}{\Psi_I(R,S)} 
\label{eq:weight}
\end{equation}
In the same spirit as the GFMC, only one of the four configurations is kept according to a heat-bath sampling among the four normalized weights $w_i/W$, with $W=\sum_{i=1}^4 w_i /4 $ being the cumulative weight. The latter is then rescaled by $W\to W\exp[-V_{SI}(R)/2+V_{SI}(R^\prime)/2-E_T]\delta\tau\}$ and associated to this new configuration for branching and computing observables. This ``plus and minus'' procedure, introduced in Ref.~\cite{Gandolfi:2014ewa} and so far only applied to systems including protons, is adopted in this work to compute the energy of PNM, as it significantly reduces the dependence of the results on $\delta\tau$.

The expectation values of observables that commute with the Hamiltonian are estimated as
\begin{align}
\langle O(\tau) \rangle =\frac{\sum_X O_T(X) W_T(X)}{\sum_X W_T(X)}\, ,
\label{eq:O_exp}
\end{align}
where 
\begin{equation}
O_T(X)=\frac{\langle \Psi_T | O | X\rangle}{\langle \Psi_T | X\rangle} \ ,\  W_T(X)=W(X) \frac{\langle \Psi_T | X\rangle}{\langle \Psi_I | X\rangle}\, .
\end{equation}
For all other observables we compute the mixed estimates
\begin{equation}
\langle O(\tau) \rangle \simeq 2 \frac{\langle \Psi_T | O | \Psi(\tau)\rangle}{\langle \Psi_T | \Psi(\tau)\rangle} - \frac{\langle \Psi_T | O | \Psi_T\rangle}{\langle \Psi_T | \Psi_T\rangle}\, ,
\label{eq:O_exp_nc}
\end{equation}
where the first and the second term correspond to the DMC and VMC expectation value, respectively.

As in standard fermion diffusion Monte Carlo algorithms, the AFDMC method suffers from the fermion sign problem. This originates from the fact that the importance-sampling wave-function is not exact and entails spuriosities from the Bosonic ground-state of the system. As a consequence, the numerator and denominator of Eq.~\eqref{eq:O_exp} are plagued by an increasing error to signal ratio for a finite sample size and large imaginary times. To alleviate the sign problem, as in Ref.~\cite{Zhang:2003zzk}, we implement an algorithm similar to the constrained-path approximation~\cite{Zhang:1996us}, but applicable to complex wave functions and propagators. The weights $w_i$ of Eq.~\eqref{eq:weight} are evaluated with
\begin{equation}
\frac{\Psi_I(R^\prime, S^\prime)}{\Psi_I(R,S)} \to {\rm Re} \left\{ \frac{\Psi_T(R^\prime, S^\prime)}{\Psi_T(R,S)} \right\}\, .
\end{equation}
and they are set to zero if the ratio is negative. Unlike the fixed-node approximation, which is applicable for scalar potentials and for cases in which a real wave function can be used, the solution obtained from the constrained propagation is not the a rigorous upper-bound to the true ground-state energy~\cite{Wiringa:2000gb}. To remove the bias associated with this procedure, the configurations obtained from a constrained propagation are further evolved using the following positive-definite importance sampling function~\cite{Pederiva:2004iz,Lonardoni:2018nob}
\begin{equation}
\Psi_G(X) = \sqrt{ { \rm Re} \{\Psi_T(X)\}^2 + \alpha\,{ \rm Im} \{\Psi_T(X)\}^2}
\end{equation}
where we typically take $\alpha=0.5$. Along this unconstrained propagation, the expectation value of the energy $E_{\rm UC}(\tau)$ is estimated according to Eq.~\eqref{eq:O_exp}. The only difference, needed to compensate for the change of the guiding wave function, is that the weights need to be rescaled as 
\begin{equation}
W(X)\to W(X) \frac{\Psi_G(X_0)}{\Psi_T(X_0)}\, ,
\end{equation}
where $X_0$ is the initial configuration of the unconstrained propagation at $\tau=\tau_0$. In a typical calculation, $\sim 400$ independent unconstrained propagations, each comprised of an average of $\sim 140,000$ configurations, are performed to control statistical fluctuations. The asymptotic value $E_0=  \lim_{\tau \to \infty} E_{\rm UC}(\tau)$ is found by fitting the imaginary-time behavior of $E_{\rm UC}(\tau)$ with a single-exponential function, as in Refs.~\cite{Pudliner:1997ck}. Since the the expectation values are substantially correlated in $\tau$, the likelihood function is computed by fully taking into account the covariance matrix of the data. We have explicitly checked that the number of independent unconstrained propagations is large enough to avoid potential instabilities arising when the covariance matrix has at least only very small eigenvalue~\cite{Jang:2011fp}. The confidence interval associated to $E_0$ is estimated as discussed in Sec.15.6 of Ref.~\cite{Press:1993}. The best value of the fit is perturbed in such a way that $\Delta \chi^2 =1$ from its minimum while varying the other fitting parameters to minimize the $\chi^2$. Since this procedure brings about an asymmetric confidence interval, in our results we report a symmetric error bar conservatively corresponding to the largest interval.  

\begin{figure}[h!]
\centering
\includegraphics[width=\columnwidth]{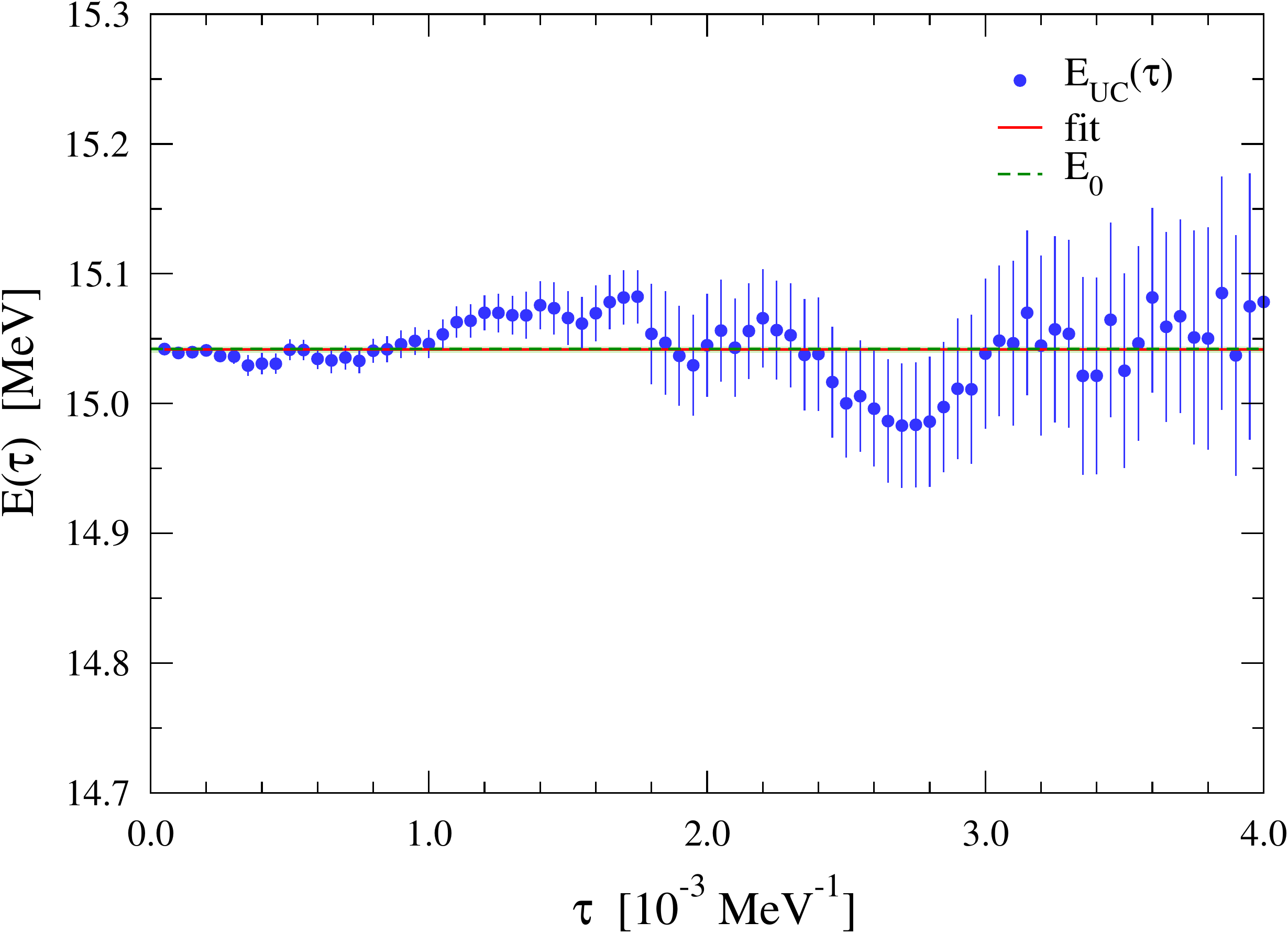}
\caption{PNM unconstrained evolution for the AV6$^\prime$ potential at $\rho=0.16$ fm$^{-3}$ for 14 neutrons in PBC. Data points (in blue) refer to $E_{\rm UC}(\tau)$ while the dashed line and the shaded (green) area indicate the asymptotic value $E_0$ with the associated uncertainty as estimated from the fit, represented by the solid (red) line.}
\label{fig:av6p_uc}
\end{figure}

\begin{figure}[h!]
\centering
\includegraphics[width=\columnwidth]{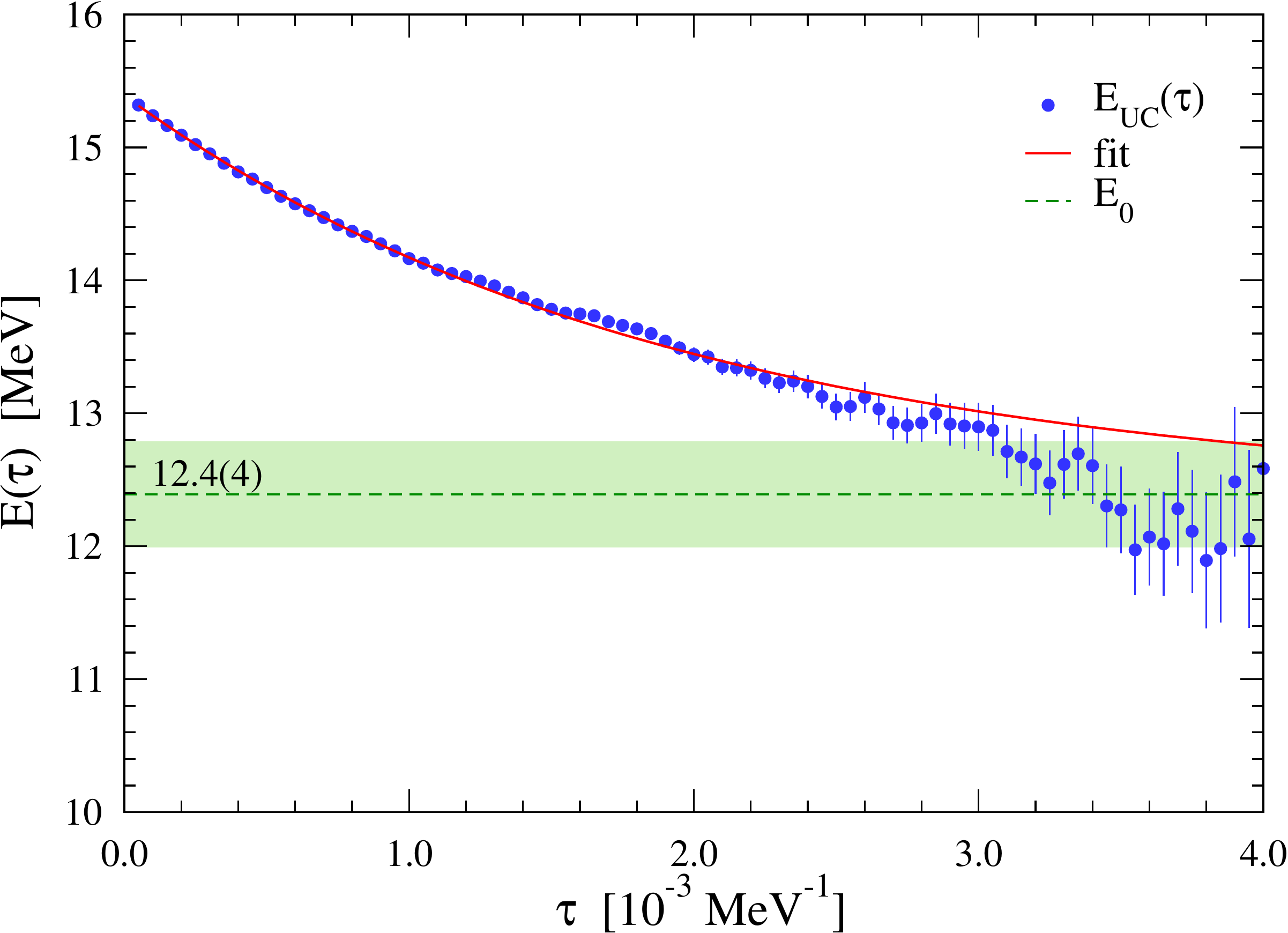}
\caption{Same as Fig.~\ref{fig:av8p_uc} for the AV8$^\prime$ interaction.}
\label{fig:av8p_uc}
\end{figure}

Unconstrained propagations have been performed in the latest AFDMC studies of atomic nuclei~\cite{Lonardoni:2017hgs,Lonardoni:2018nob,Lonardoni:2018sqo}, even though a relatively simpler fitting procedure was employed to determine the asymptotic $E_0$ and its error. On the other hand, the accuracy of the constrained approximation for neutron systems has been generally acknowledged, even in the state-of-the art AFDMC neutron-matter calculations with local chiral interactions~\cite{Lovato:2010ef,Gezerlis:2013ipa,Tews:2015ufa}. Fig.~\ref{fig:av6p_uc} indeed shows that for the AV6$^\prime$ potential at $\rho=\rho_0$ releasing the constraint brings about minor changes to the constrained results. The situation is drastically different for NN potentials that include spin-orbit terms. The unconstrained propagation for the AV8$^\prime$ potential at $\rho=0.16$ fm$^{-3}$, displayed in Fig.~\ref{fig:av8p_uc}, exhibits a a clear exponentially-decaying behavior, lowering the energy per particle by as much as $\sim 3$ MeV. We checked that including linearized spin-dependent correlations in the trial wave function yields only $\sim 0.3$ MeV of additional binding in the constrained propagation. Given the increase in the computational cost of the calculation and the need of large statistics to reliably perform the imaginary-time extrapolation, we have decided to stick to the simple central Jastrow ansatz. On the other hand, the spin-dependent backflow correlations of Ref.~\cite{Brualla:2003gw} seems to be more effective:  preliminary calculations indicate that the constrained results can be lowered by more than $\sim 1$ MeV per particle. For both AV6$^\prime$ and AV8$^\prime$, we simulated PNM using 14 neutrons in PBC, correcting for the tails of the potential and Jastrow correlations. The dependency on the box size of the AV8$^\prime$ results has been tested performing additional calculation with 38 neutrons in a PBC. It turns out that $E_{\rm UC}(\tau) - E_{\rm UC}(\tau_0)$ obtained with the two simulation boxes are fully compatible within statistical errors. Our findings for the AV8$^\prime$ interaction are consistent with the GFMC results of Ref.~\cite{Carlson:2003wm} and with the discrepancies in the spin-orbit splitting of neutron drops between AFDMC and GFMC calculations~\cite{Pederiva:2004iz}. 

\begin{figure}[h!]
\centering
\includegraphics[width=\columnwidth]{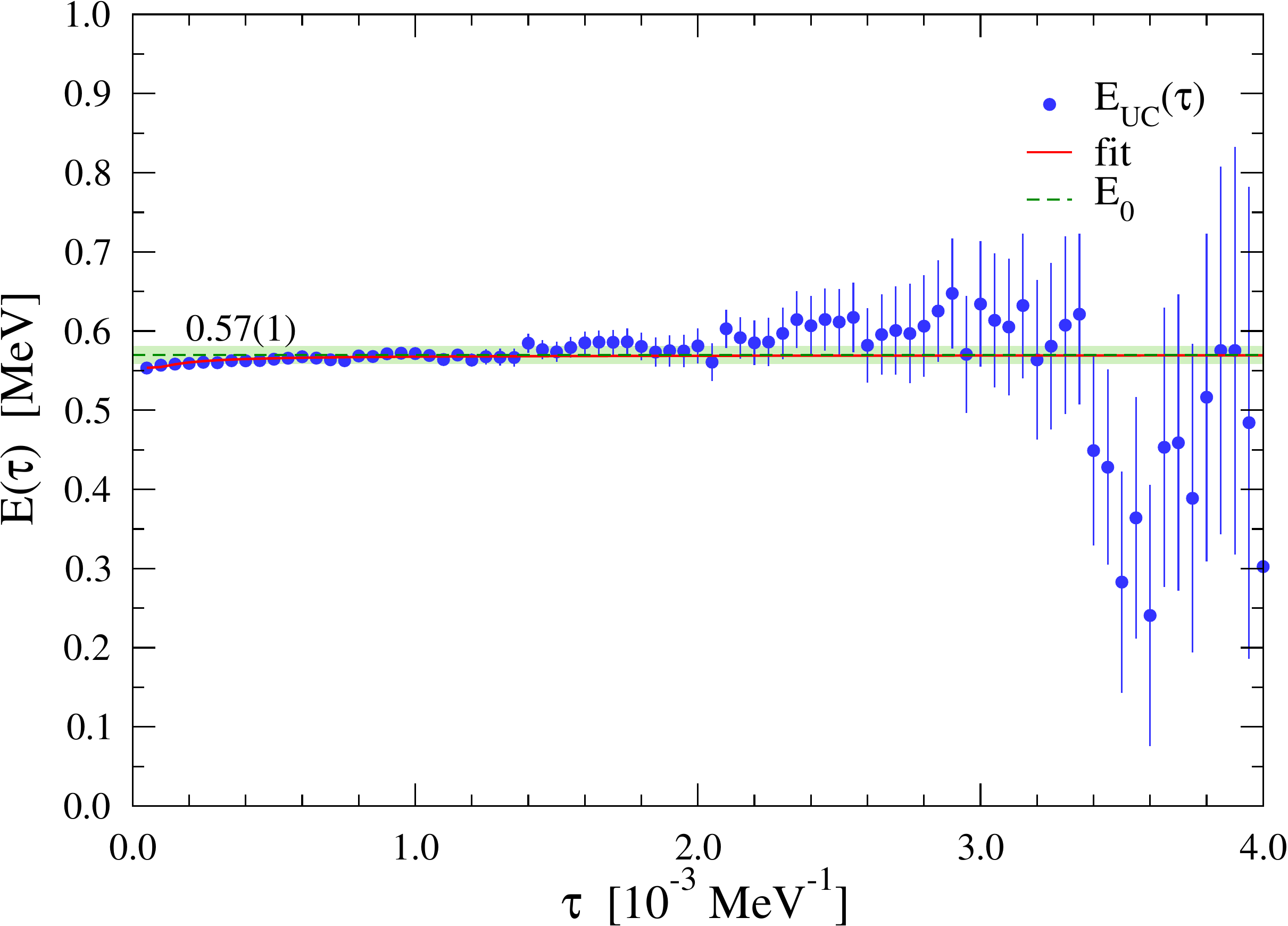}
\caption{PNM unconstrained evolution of $\langle v_{18} - v_{8}^\prime\rangle$ for at $\rho=0.16$ fm$^{-3}$ for 14 neutrons in PBC. The notation is the same as in Fig.~\ref{fig:av8p_uc}.}
\label{fig:v18_v8}
\end{figure}

Analogously to the GFMC method, when computing the full AV18 and NV2 two-body interactions, the propagation is performed with the simplified $v_8^\prime$ potential, described in Section \ref{sec:nuc_int}. The expectation value $\langle v_{18} - v_{8}^\prime\rangle$ is evaluated in perturbation theory according to Eq.~\eqref{eq:O_exp_nc}. As shown in Fig.~\ref{fig:v18_v8} for $\rho=0.16$ and 14 neutrons with PBC, the potential energy difference remains fairly stable during the unconstrained propagation. We fit its imaginary time behavior with a simple inverse polynomial formula with up to $1/\tau^2$ powers and estimate the error on the asymptotic value accordingly.

\section{Results}
\label{sec:results}
We compare the PNM equation of state as obtained from the three independent many-body methods described in Section~\ref{sec:mbm}, using the Argonne and the Norfolk families of NN interactions. As for the AFDMC, we present results corresponding to both the constrained (AFDMC-CP) and unconstrained (AFDMC-UC) imaginary-time propagations. To minimize finite-size effects, AFDMC-CP calculations are carried out with 66 neutrons in a box with PBC. On the other hand, the unconstrained energy is estimated by adding to the AFDMC-CP values the energy difference $E_{\rm UC}(\tau) - E_{\rm UC}(\tau_0)$ computed simulating 14 neutrons with PBC. This procedure significantly reduces the computational cost of the calculation. Its accuracy is validated by the successful comparison of unconstrained propagations with 14 and 38 neutrons with PBC, discussed in the previous Section. 

\begin{figure}[!h]
\centering
\includegraphics[width=\columnwidth]{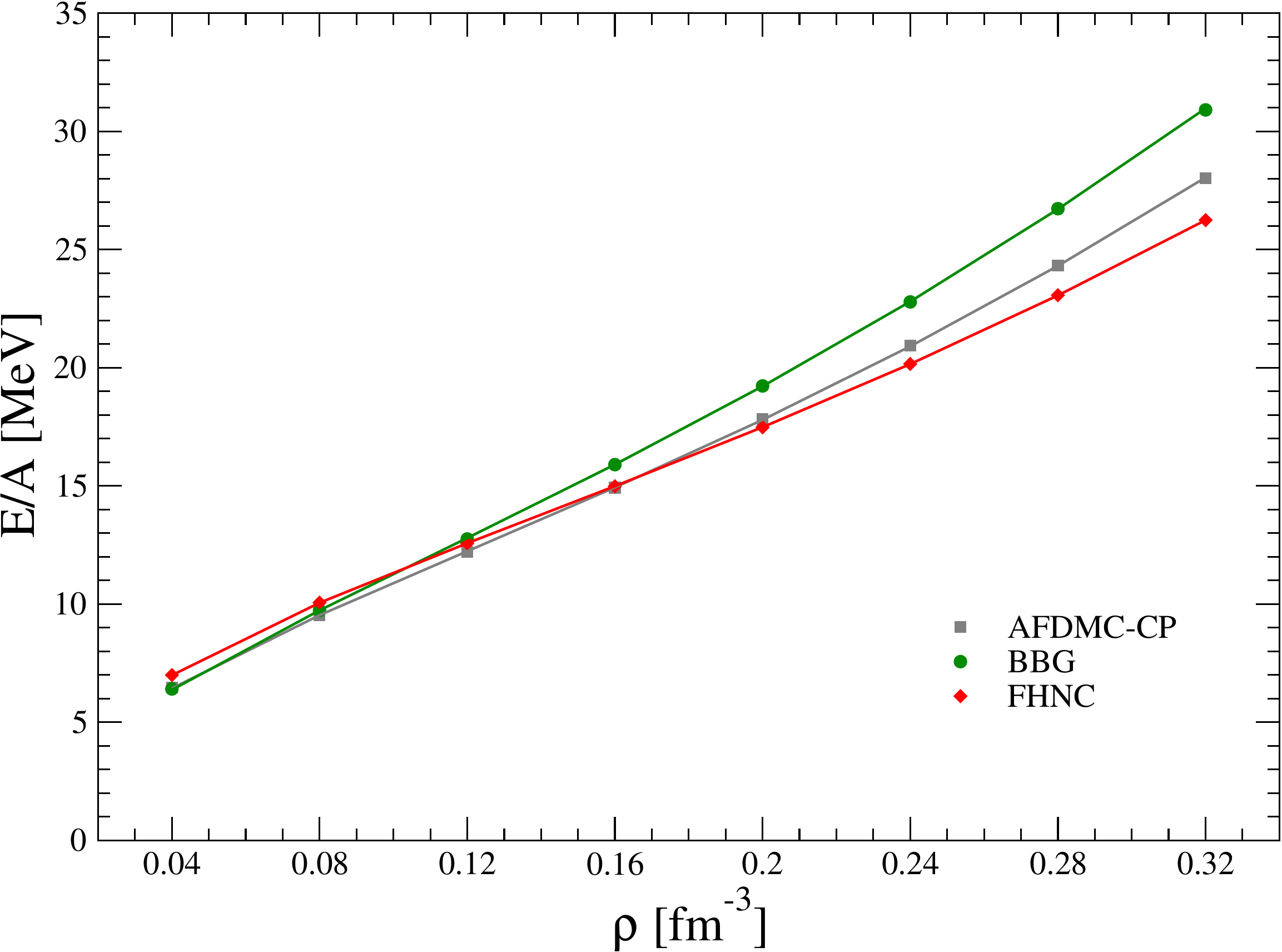}
\includegraphics[width=\columnwidth]{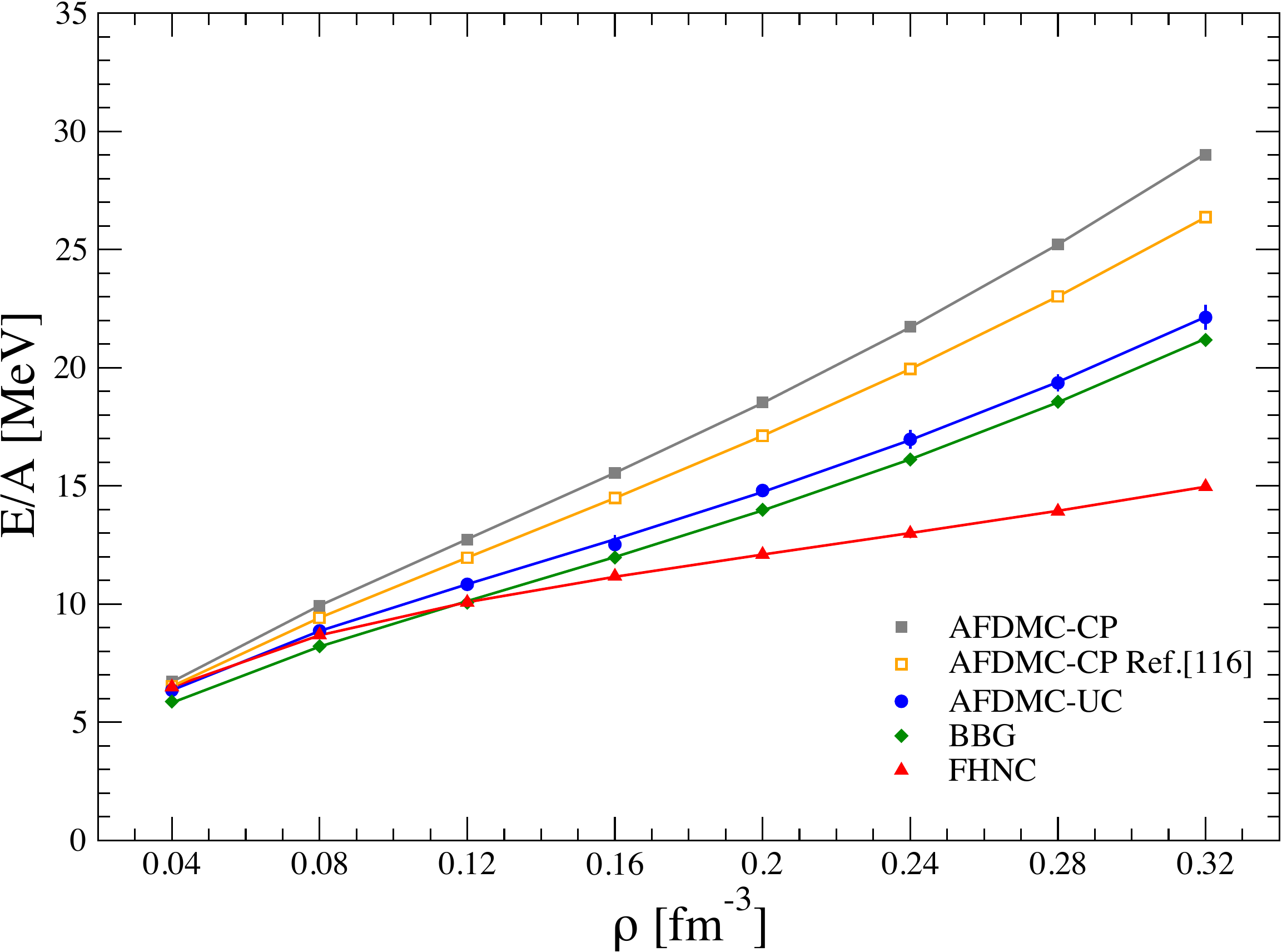}
\includegraphics[width=\columnwidth]{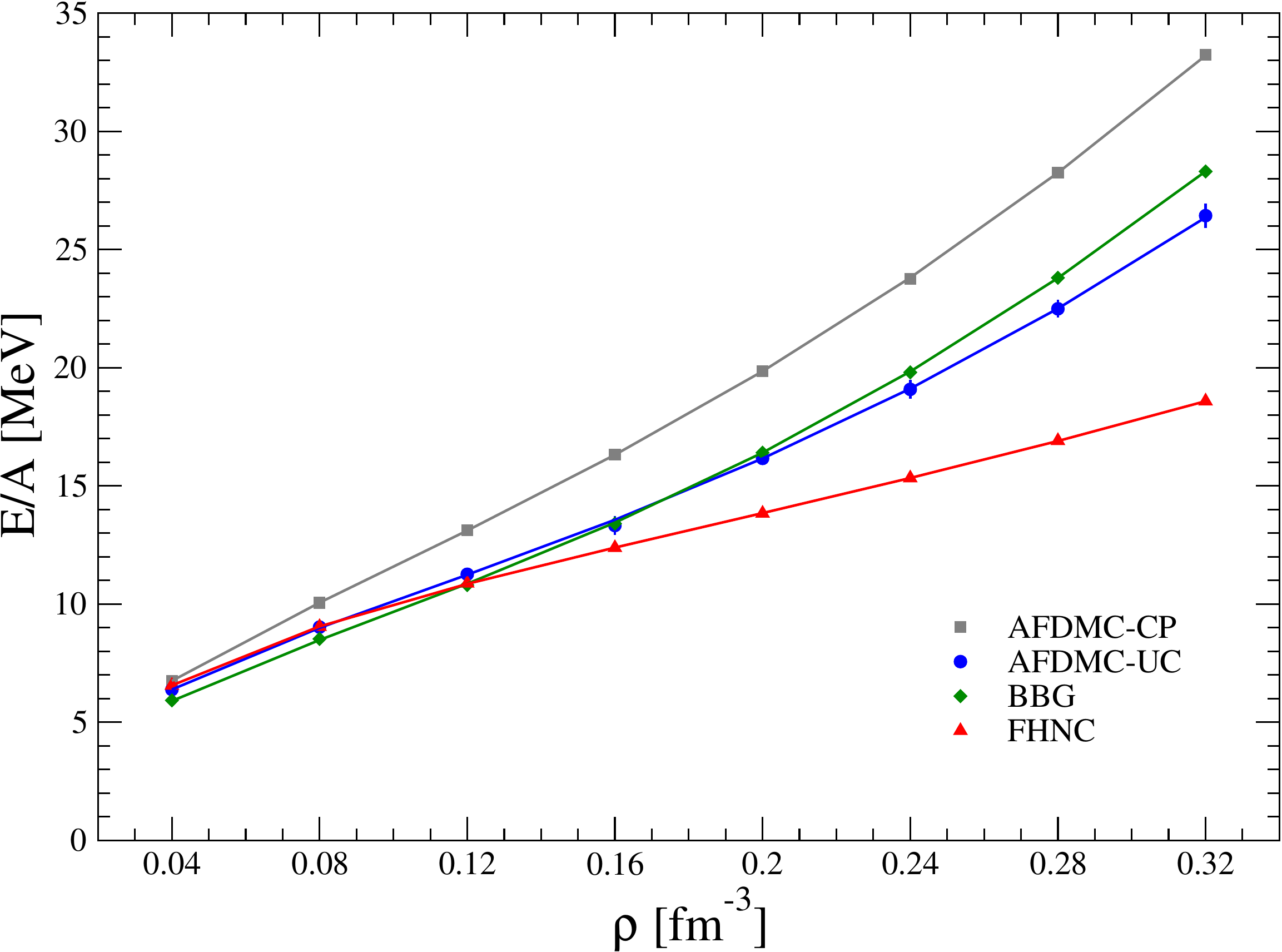}
\caption{(Color online) Energy per particle of PNM as a function of density calculated with the BBG (green diamonds), FHNC/SOC (red triangles), AFDMC-CP (grey squares) and AFDMC-UC (solid blue points) many-body approaches. Results for the AV6$^\prime$, AV8$^\prime$, and AV18 potentials are shown in the upper, middle, and lower panels, respectively. The curves correspond to the polynomial fit of Eq.~\eqref{eq:pol_fit}}
\label{fig:PNM_AV}
\end{figure}

\begin{figure*}[t]
\centering
\includegraphics[width=\columnwidth]{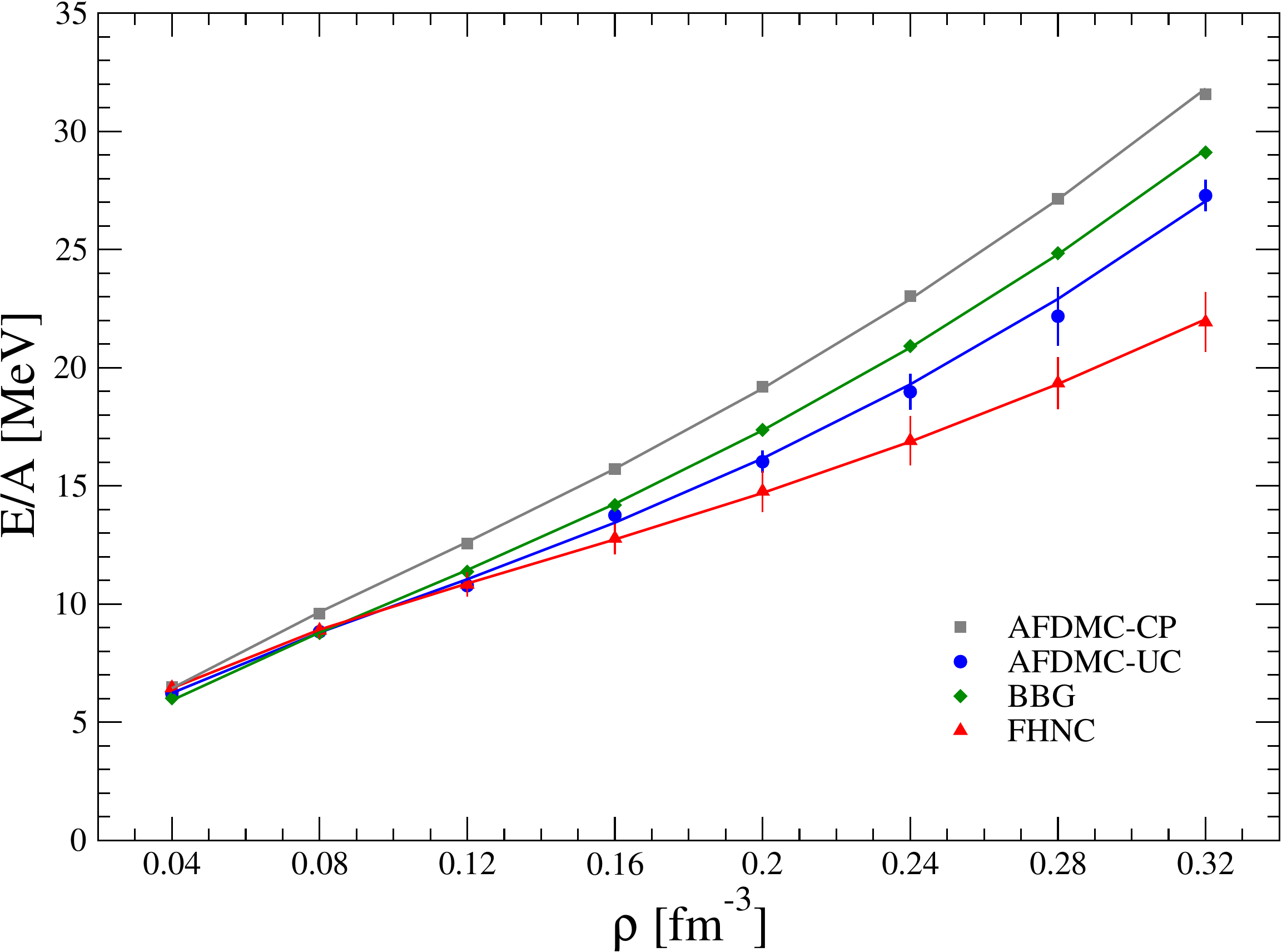}
\includegraphics[width=\columnwidth]{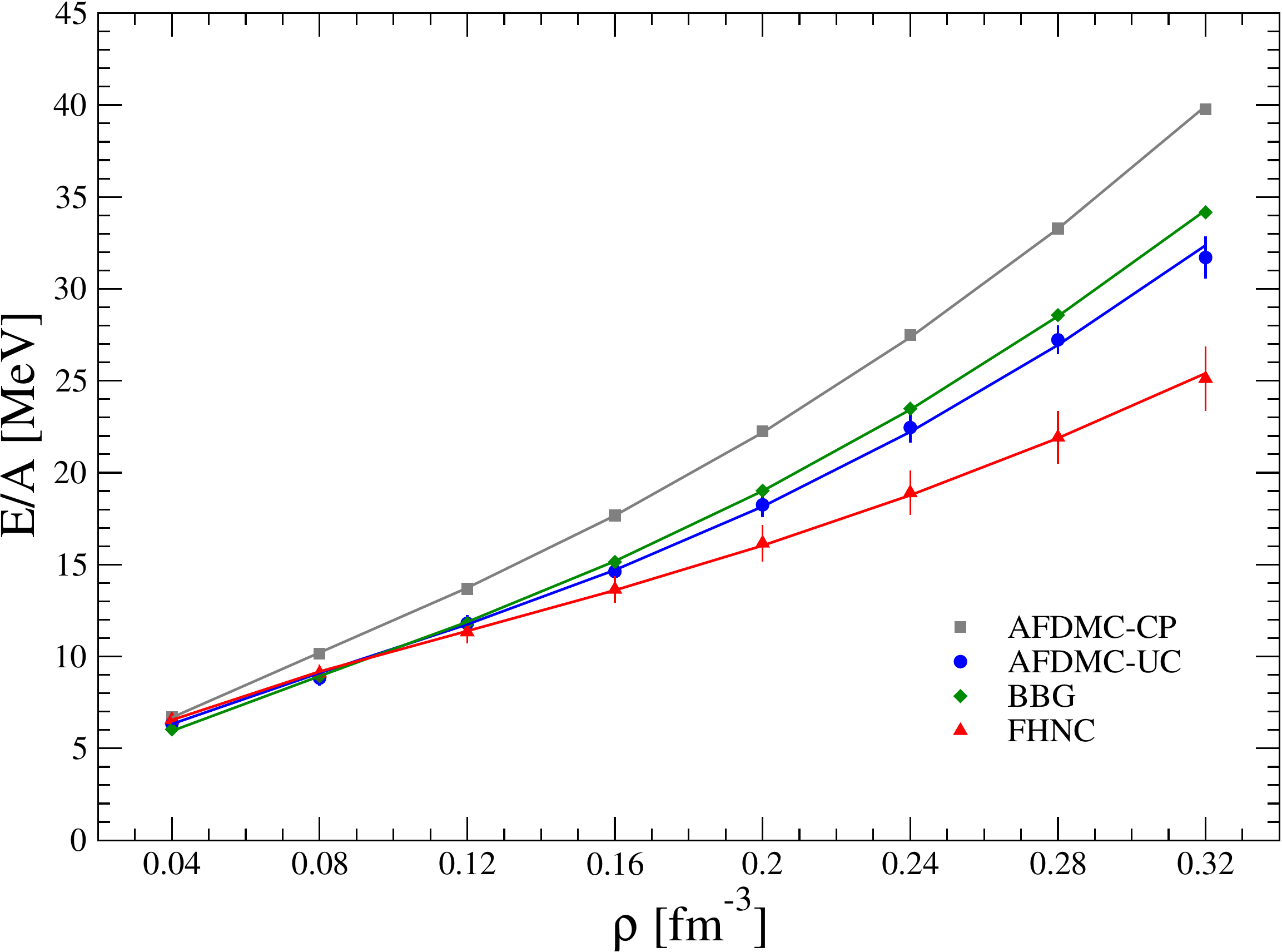}
\includegraphics[width=\columnwidth]{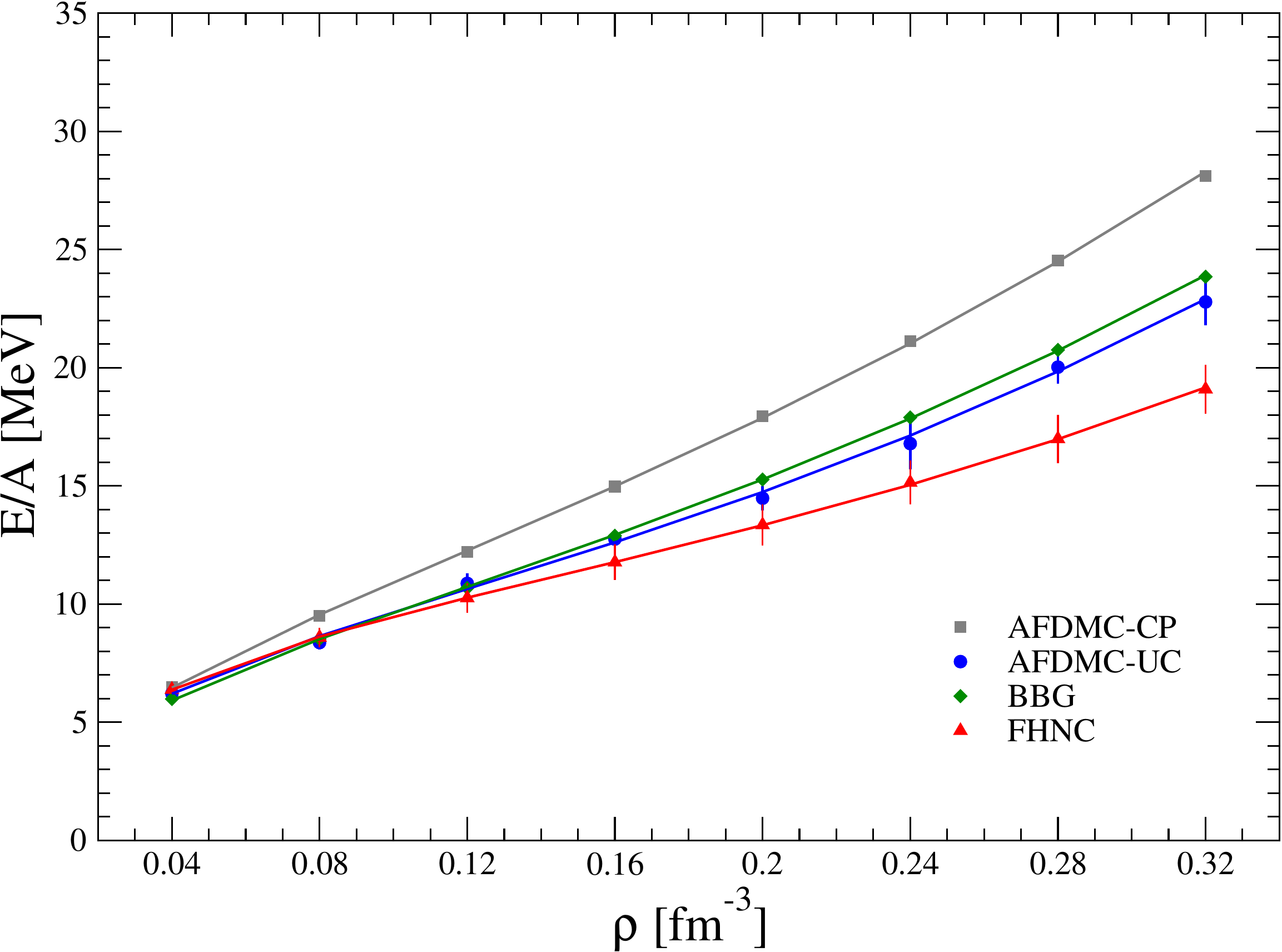}
\includegraphics[width=\columnwidth]{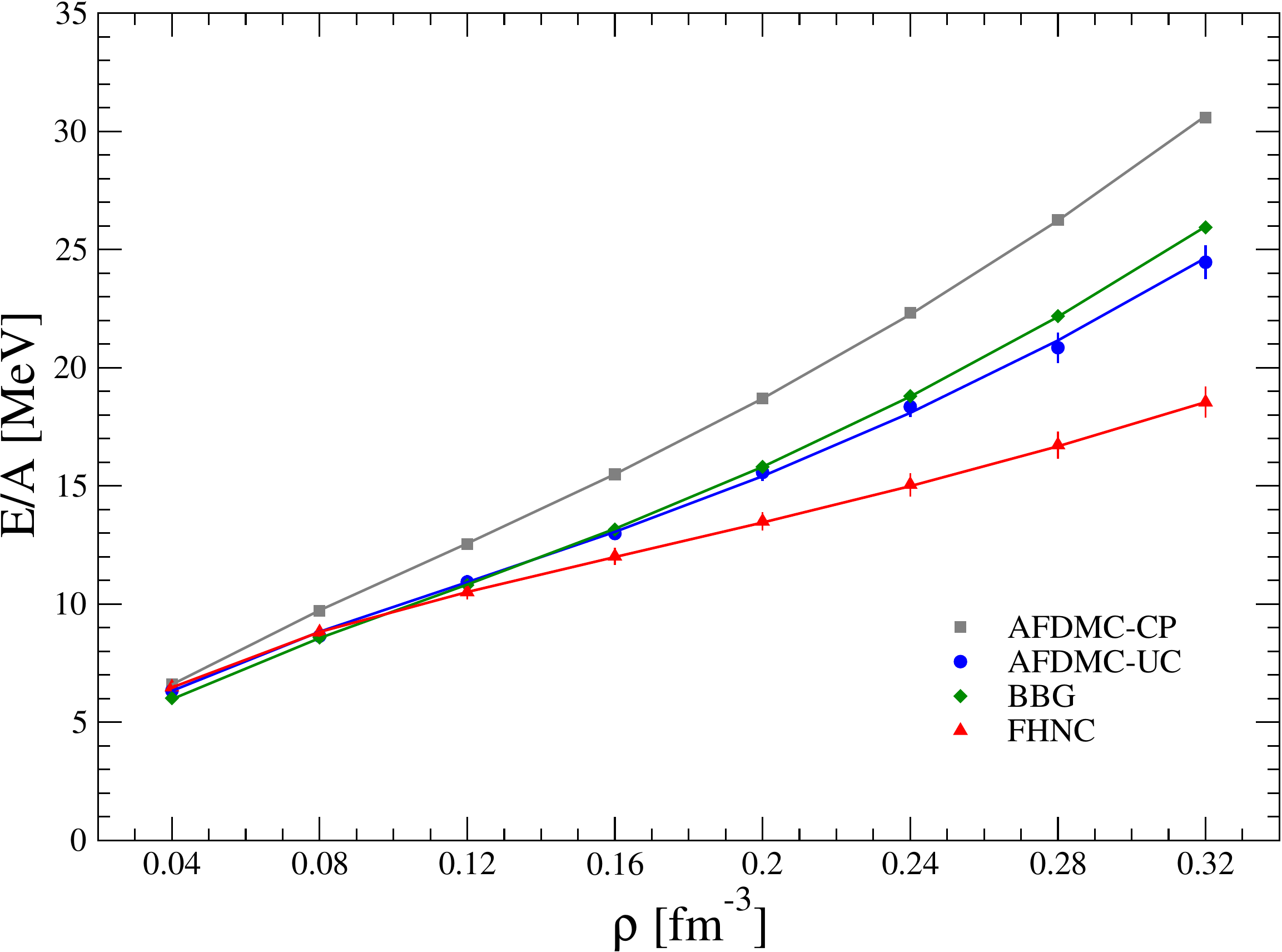}
\caption{Same as Fig.~\ref{fig:PNM_AV} for the NV2-Ia (upper left panel), NV2-Ib (upper right panel), NV2-IIa (lower left panel), and NV2-IIb (lower right panel) potentials.}
\label{fig:PNM_NV2}
\end{figure*}

In the upper, medium, and lower panels of Fig.~\ref{fig:PNM_AV} we show the PNM equation of state for the AV6$^\prime$, AV8$^\prime$, and AV18 potentials, respectively. The curves in the plot correspond to the following polynomial fit for the density dependence of the energy per particle
\begin{equation}
\frac{E(\rho)}{A} = a_{2/3} \left(\frac{\rho}{\rho_0}\right)^{2/3} + a_{1} \left(\frac{\rho}{\rho_0}\right) + a_{2} \left(\frac{\rho}{\rho_0}\right)^2\, ,
\label{eq:pol_fit}
\end{equation}
where $\rho=0.16$ fm$^{-3}$ is the nuclear saturation density. The first term corresponds to a free Fermi gas, while the second and third are inspired by the cluster expansion of the energy expectation value, truncated at the three-body level. We have checked that the four-parameter fitting function of Ref.~\cite{Gandolfi:2013baa} produces overlapping curves to the one obtained from Eq.~\eqref{eq:pol_fit}.

Consistently with Ref.~\cite{Baldo:2012nh}, when the AV6$^\prime$ interaction is employed, the three many-body methods provide similar results for $E(\rho)/A$. Generally, BBG yields the most repulsive EoS, FHNC/SOC the softest, and the AFDMC-CP values are in between the other two. Even at $\rho=2\rho_0$, the maximum spread among the different methods remains within $\sim 5$ MeV per particle. Note that the FHNC/SOC calculations shown in this work are more sophisticated than those of Ref.~\cite{Baldo:2012nh}, as more elementary diagrams -- at and beyond the FHNC/4 approximation -- are accounted for. This leads to accurate estimates for the energy per particle, particularly when spin-orbit correlations are not included. The AFDMC-UC energies for AV6$^\prime$ are not shown, as, within error bars, they overlap with the AFDMC-CP ones -- see the unconstrained propagation of Fig.~\ref{fig:av6p_uc}.

The inclusion of spin-orbit components of the AV8$^\prime$ potential brings about an overall attraction in PNM with respect to the AV6$^\prime$ results, for both BBG and FHNC/SOC methods. This appears to be a consequence of the isospin asymmetry: in GFMC calculations for light nuclei, AV6$^\prime$ is more attractive for isospin-symmetric nuclei, but AV8$^\prime$ is more attractive in neutron-rich systems~\cite{wiringa02}.  
For example, as seen in the associated force evolution table~\cite{evolution}, the two models give the same energy for $^6$He, while in $^8$He, AV8$^\prime$ is $\simeq1$ MeV more bound.
Also, the difference in binding between $^8$Be and $^8$He is $26.9$ MeV for AV6$^\prime$ and $24.2$ MeV for AV8$^\prime$, implying that AV8$^\prime$ is bringing in relatively more attraction for the neutron-rich systems.

On the other hand, the AFDMC-CP energies per particle for AV8$^\prime$ are slightly larger than those obtained with AV6$^\prime$, and they lie well above both BHF and FHNC/SOC results, already at relatively small densities. At $\rho=\rho_0$, BBG, FHNC/SOC, and AFDMC-CP provide $11.97$ MeV, $11.2(2)$ MeV, and $15.55(1)$ MeV per particle, respectively. The unconstrained propagation significantly lowers the AFDMC-CP estimates, bringing them in much better agreement with BBG calculations. At $\rho=\rho_0$, the AFDMC-UC value turns out to be $12.5(3)$ MeV, while at $\rho=2\rho_0$ the unconstrained propagation yields $22.1(5)$ MeV, to be compared to the $29.01(1)$ MeV of the constrained approximation. The curve corresponding to the AFDMC-CP calculations of Ref.~\cite{Gandolfi:2013baa} lies below the AFDMC-CP obtained with the ``plus and minus'' importance-sampling algorithm. The differences between the two constrained approximations are largely due to the dependence on the central Jastrow correlations of the importance-sampling algorithm utilized in Ref.~\cite{Gandolfi:2013baa}. As noted in Ref.~\cite{Gezerlis:2013ipa}, for the local N2LO $\chi$EFT potential this unphysical dependence on the Jastrow function can be as large as $0.6$ MeV per particle already at $\rho=0.1$ fm$^{-3}$. Note that this dependence is completed removed once the ``plus and minus'' procedure is employed.

The FHNC/SOC results stay well below both the BBG and AFDMC-UC ones, with the spread increasing with the density. This behavior is most likely due to the oversimplified treatment of spin-orbit correlations, which become less accurate at higher densities, as contributions arising from clusters involving more than three nucleons cannot be neglected. We explicitly checked that, as pointed out in Refs.~\cite{Lovato:2011ij,Lovato:2010ef,Baldo:2012nh}, when the spin-obit correlations are turned off, FHNC/SOC and AFDMC-CP are in much better agreement. 

It is remarkable that the AFDMC-UC and BBG predictions are quite similar, the differences remaining well below $1$ MeV per particle up to $\rho=2\rho_0$. This corroborates the accuracy of the extrapolation of the unconstrained energy. From a different point of view, the good agreement between the AFDMC-UC and the BBG results can be interpreted as an indication of the accuracy of the BHF approximation of the BBG hole-line expansion, thus indirectly confirming the smallness of the contribution of the three-hole-diagrams to the energy per nucleon of PNM~\cite{baldo00, Jia18} within the continuous choice of the single particle potential of Eq. \eqref{spp}.

As in light nuclei, the AV18 potential, is more repulsive than AV8$^\prime$ for all the many-body methods considered in this work. In particular, as the density increases, the differences in partial waves higher than $P$ become more and more important. The AFDMC-CP results turn out to be biased to a slightly smaller extent than for AV8$^\prime$: the unconstrained propagation at $\rho=\rho_0$ and $\rho=2\rho_0$ lowers the energy per particle by $\sim 2.2$ MeV and $\sim 5.3$ MeV, respectively. The BBG and AFDMC-UC predictions are again very close: the maximum difference remains below $\sim 1$ MeV per particle. Contrary to the AV8$^\prime$ case, AFDMC-UC yields slightly less repulsion than BBG. At $\rho > \rho_0$, the FHNC/SOC results lie significantly below those computed within both AFDMC-UC and BBG. This might once more be ascribed to the three-body truncation in the cluster expansion of the spin-orbit correlations. 

Fig.~\ref{fig:PNM_NV2} displays the energy per particle of the NV2-Ia, NV2-Ib, NV2-IIa, and  NV2-IIb potentials as computed within the different many-body methods and their polynomial fit using the expression of Eq.~\eqref{eq:pol_fit}. The picture that emerges is largely consistent with the one already discussed for the AV8$^\prime$ and AV18 interactions. The AFDMC-CP calculations suffers from a sizable systematic error that increases with the density. Releasing the constraint in the imaginary-time propagation lowers the energy per particle by as much as $\sim 8$ MeV at $\rho=2\rho_0$ for the NV2-Ib model. The good agreement between BBG and AFDMC-UC is to a large extent confirmed, as the discrepancies between the two methods are smaller than $\sim 2.5$ MeV for all the densities and potentials we analyzed. Once again, for densities larger than $\rho_0$, FHNC/SOC calculations yield considerably lower energies than  BBG and AFDMC-UC.

By taking the AFDMC-UC results as references, comparing the EoS obtained using the AV18 and the NV2 potentials we observe that, with the exception of the NV2-Ib case, the maximum spread among the curves is well within $5$ MeV per particle up to $\rho=2\rho_0$. In fact, for densities smaller than nuclear saturation, the differences are always below $\sim 1$ MeV per particle. It has be noted that the NV2-Ib interaction has a relatively hard regulator ($R_S=0.7$ fm) and has been fitted against the smallest energy range of NN scattering data, up to $E_{\rm lab}= 125$ MeV. From Fig.~\ref{fig:ps_chiral}, the P-wave phase shifts computed with the NV2-Ib significantly deviate from the experimentally-extracted ones already for $E_{\rm lab}\lesssim 200$ MeV, corresponding to densities smaller than nuclear saturation. Hence, fitting NN scattering up to higher energies seems to be rather effective in controlling their predictions for the energy per particle of infinite neutron matter up to relatively-high densities.

\section{Conclusions}
\label{sec:conclusions}
We have carried out benchmark calculations of the energy per particle of pure neutron matter as a function of the baryon density, employing two distinct families of coordinate-space nucleon-nucleon potentials in three independent nuclear many-body methods: the AFDMC, the FHNC/SOC, and the BBG. As for the nuclear Hamiltonians, we have considered the phenomenological Argonne AV6$^\prime$, AV8$^\prime$~\cite{wiringa02}, and AV18 two-body interactions~\cite{Wiringa:1994wb}, and the set of Norfolk $\chi$EFT NV2 potentials~\cite{Piarulli:2014bda,Piarulli:2016vel}, which explicitly includes the $\Delta$ isobar intermediate state. With the exception of AV6$^\prime$, these potentials are characterized by relatively strong spin-orbit components, needed to reproduce the NN phase shifts in $P$ and higher odd partial waves. 

Our pure neutron matter AFDMC calculations are performed using the ``plus and minus'' importance-sampling algorithm, introduced in Ref.~\cite{Gandolfi:2014ewa} to treat atomic nuclei and isospin-symmetric and asymmetric nuclear matter. On the other hand, previous application of the AFDMC method to purely-neutron systems used a different importance sampling for both the spacial coordinates and the auxiliary fields. Extending the analysis of Refs.~\cite{Carlson:2003wm,Pederiva:2004iz}, we have investigated the systematic error of the AFDMC method arising from constraining the imaginary-time propagation to alleviate the fermion-sign problem. We have performed unconstrained imaginary-time propagations up to $0.004$ MeV$^{-1}$, extrapolating the asymptotic value for the energy per particle using a single-exponential fit. By computing the covariance matrix of the data to account for the correlations among the AFDMC samples, we are able to estimate the uncertainty of the asymptotic energy by varying the $\chi^2$ contour of the fit. The FHNC/SOC method has been improved by systematically including sets of elementary diagrams, at and beyond the FHNC/4 approximation, through the use of three-point superbonds in the diagrammatic expansion. 

When the AV6$^\prime$ interaction is employed, AFDMC, FHNC/SOC, and BBG yield similar energies per particle, the maximum difference among the methods remaining smaller than $5$ MeV per particle up to $\rho=2\rho_0$. The excellent agreement between AFDMC and FHNC/SOC calculations has to be ascribed to both the improved sampling in the AFDMC method and to the inclusion of the elementary diagrams in FHNC/SOC. Releasing the constraint on the imaginary-time propagation does not bring about appreciable difference with respect to the AFDMC-CP results. On the other hand, when spin-orbit terms are present in the nuclear interaction, we find that performing unconstrained propagations is crucial to reliably compute the equation of state of neutron matter. Simple constrained propagations significantly overestimate the energy per particle, with the bias increasing with the density. For instance, when the AV18 potential is used, the difference between AFDMC-CP and  AFDMC-UC calculations can be as large as $\sim 3$ MeV at $\rho=\rho_0$ and $\sim 7$ MeV at $\rho=\rho_0$. Similar trends are also found for the AV8$^\prime$ potential and all NV2 interactions and we can reasonably expect that analogous systematic errors affect the AFDMC calculations of neutron-matter properties carried out with local N2LO $\chi$EFT Hamiltonians~\cite{Gezerlis:2013ipa,Gezerlis:2014zia,Lynn:2015jua,Tews:2015ufa,Tews:2018kmu}. 

The AFDMC-UC predictions are in good agreement with those of the BBG approach. For both AV18 and the NV2 potentials, the discrepancies between the two methods remain well within $3$ MeV per particle, with the AFDMC-UC method always providing less repulsion than the BBG. This highly non trivial outcome of our comparison has been enabled by the possibility of performing unconstrained propagations in AFDMC. As a matter of fact, the AFDMC-CP equations of state are sizably above both BBG and AFDMC-UC ones. The FHNC/SOC energies per particle are consistently below those computed within the other two many-body  methods, particularly for densities larger than $\rho_0$. This is likely to be ascribed to the somewhat oversimplified treatment of spin-orbit correlations, whose contributions are only retained up to the three-body cluster level. Limiting our analysis to $\rho \le \rho_0$, where higher-order terms in the cluster expansion are smaller, FHNC/SOC and AFDMC-UC agree up to $1$ MeV per particle, while the difference between AFDMC-CP and FHNC/SOC turns out to be significantly larger. 

The AV18 potential fits NN scattering data with $\chi^2\sim 1$ in the energy range $0\leq E_{\rm lab} \leq 350$ MeV, while NV2 potentials are constrained up to lower energies: $0\leq E_{\rm lab} \leq 125$ and $0\leq E_{\rm lab} \leq 200$ MeV for class I and class II, respectively. Hence, the AV18, NV2-IIa, and NV2-IIb reproduce the experimental proton-neutron scattering phase shifts in the $^1$S$_0$, $^3$P$_0$, $^3$P$_1$, and $^3$P$_2$ partial waves to higher energies than NV2-Ia and NV2-Ib. Since in highly-degenerate matter neutron-neutron collisions mostly take place in the vicinity of the Fermi surface, one can reasonably expect that potential models capable of reproducing NN scattering to higher $E_{\rm lab}$ will more reliably predict the EoS at larger densities. Our AFDMC-UC calculations indicate that this is indeed the case. The maximum spread among the energies per particle obtained using the AV18, NV2-IIa, and NV2-IIb potentials is well within $4$ MeV per particle up to twice nuclear saturation density. On the other hand, including NV2-Ia and NV2-Ib, the spread among the  models can be as large as $\sim 9$ MeV per particle. 

This work extends the benchmark calculations carried out in the literature~\cite{Baldo:2003yp,Bombaci:2004iu,Baldo:2012nh} and it is not aimed at obtaining a realistic description of the neutron matter EoS, for which three-body forces are required. Two classes of $\chi$EFT three-nucleon interactions consistent with the $\Delta$-full NN potentials employed in this work have been derived and successfully applied to describe the spectrum of light nuclei~\cite{Piarulli:2017dwd} and the $\beta$-decay of $^3$H~\cite{Baroni:2018fdn}. Once implemented in our many-body methods, we will compute the EoS and check their compatibility with astrophysical constraints, gauging potential regulator artifacts~\cite{Lovato:2011ij,Tews:2015ufa,Huth:2017wzw} and the convergence of the chiral expansion. 

\section{Acknowledgements}
We thank J. Carlson, D. Lonardoni, L. Riz, and I. Tews for valuable discussions. This research is supported by the U.S. Department of Energy, Office of Science, Office of Nuclear Physics, under contracts DE-AC02-06CH11357 (A. L. and R. B. W.) and under the FRIB Theory Alliance award DE-SC0013617 (M. P.). Under an award of computer time provided by the INCITE program, this research used resources of the Argonne Leadership Computing Facility at Argonne National Laboratory, which is supported by the Office of Science of the U.S. Department of Energy under contract DE-AC02-06CH11357. Numerical calculations have been made possible also through a CINECA-INFN  agreement, providing access to resources on MARCONI at CINECA.  

\bibliography{biblio}

\end{document}